\documentclass{tep}
\pdfoutput=1 
\usepackage{topwidth}

\begin{document}

\thispagestyle{empty}
\def\thefootnote{\fnsymbol{footnote}}
\setcounter{footnote}{1}
\null
\mbox{}\hfill  
FR-PHENO-2015-006, ZU-TH 26/15  \\
\vskip 0cm
\vfill
\begin{center}
  {\Large \boldmath{\bf 
      Techniques for the treatment of IR divergences \\
      in decay processes at NLO \\
      and application to the top-quark decay
    }
    \par} \vskip 2.5em
  {\large
    {\sc Lorenzo Basso$^{1,2,3}$, Stefan Dittmaier$^{1}$, \\[.3em]
      Alexander Huss$^{1,4,5}$ and Luisa Oggero$^{1}$
    }\\[1ex]
    {\small \it 
      $^1$ Albert-Ludwigs-Universit\"at Freiburg, Physikalisches Institut, 
      D-79104 Freiburg, Germany \\
      $^2$ Universit\'e de Strasbourg, IPHC, 23 rue du Loess 67037 Strasbourg,
      France \\
      $^3$ CNRS, UMR7178, 67037 Strasbourg, France \\
      $^4$ Institute for Theoretical Physics, ETH, CH-8093 Z\"urich, Switzerland \\
      $^5$ Department of Physics, University of Z\"urich, CH-8057 Z\"urich, Switzerland
    }
    \\[2ex]
  }
  \par \vskip 1em
\end{center}\par
\vskip .0cm \vfill {\bf Abstract:} \par
We present the extension of two general algorithms for the treatment of infrared singularities arising in electroweak corrections to decay processes at next-to-leading order: the dipole subtraction formalism and the one-cutoff slicing method.
The former is extended to the case of decay kinematics which has not been considered in the literature so far. The latter is generalized to production and decay processes 
with more than two charged particles, where new ``surface'' terms arise.
Arbitrary patterns of massive and massless external particles are considered, 
including the treatment of infrared singularities in dimensional or mass regularization.
As an application of the two techniques we present the calculation of the next-to-leading-order 
QCD and electroweak corrections to the top-quark decay width including all off-shell and
decay effects of intermediate $\PW$ bosons.
The result, e.g., 
represents a building block of a future calculation of NLO 
electroweak effects to off-shell top-quark pair ($\PWp\PWm\Pqb\Paqb$) production.
Moreover, this calculation can serve as the first step towards an 
event generator for top-quark decays at next-to-leading order accuracy, which can be used to attach top-quark decays to complicated many-particle top-quark processes, such as for $\Pqt\Paqt+\PH$ or $\Pqt\Paqt+\text{jets}$.
\par
\vskip 1cm
\noindent
July 2015
\par
\null
\setcounter{page}{0}
\clearpage
\def\thefootnote{\arabic{footnote}}
\setcounter{footnote}{0}



\section{Introduction}
\label{sec:intro}

The calculation of radiative corrections to cross-section predictions at next-to-leading order (NLO) saw tremendous progress in the past decades leading to a high degree of automation of calculations.
This advancement was also driven by the development of general techniques for the treatment of infrared (IR) singularities.
Such divergences arise in the intermediate steps of the calculation involving massless states and cancel in the final result by virtue of the 
Kinoshita--Lee--Nauenberg (KLN) and factorization theorems.
The various methods that were proposed in the literature can be broadly classified into two types: 
the phase-space-slicing~\cite{Giele:1991vf,Giele:1993dj,Baur:1998kt,Keller:1998tf,Harris:2001sx} 
and the subtraction formalisms~\cite{Ellis:1980wv,Ellis:1989vm,Mangano:1991jk,Kunszt:1992tn,Frixione:1995ms,Catani:1996jh,Catani:1996vz,Nagy:1996bz,Frixione:1997np,Campbell:1998nn,Dittmaier:1999mb,Phaf:2001gc,Catani:2002hc,Dittmaier:2008md,Chung:2010fx}.
Both approaches allow for the analytic cancellation of all IR singularities without jeopardising the flexibility of a fully differential numerical phase-space integration.

In this part of the 
paper we consider the so-called \emph{dipole subtraction formalism},
which is one of the most established subtraction methods.
It was continuously extended since its original formulation in 
Ref.~\cite{Catani:1996vz}, which was restricted to the computation of NLO QCD radiative corrections involving only massless partons, to cover all possible cases that are relevant in scattering processes.
In Ref.~\cite{Dittmaier:1999mb} the dipole formalism was developed for the calculation of NLO QED corrections where the emission of photons from both massless and massive fermions was considered.
The extension of the original algorithm for NLO QCD calculations to the case with massive partons 
was worked out in Refs.~\cite{Phaf:2001gc,Catani:2002hc}.
The treatment of non-collinear-safe observables in QED corrections was presented in Ref.~\cite{Dittmaier:2008md} where the formalism was further extended to cover the various fermion--photon splittings not considered in Ref.~\cite{Dittmaier:1999mb}.

For the case of decay processes, subtraction terms were constructed in Ref.~\cite{Melnikov:2011qx} for the calculation of NLO QCD corrections to the radiative decay of the top-quark. 
These subtraction terms were based on the results of Ref.~\cite{Campbell:2004ch} and constructed 
to fit into the dipole subtraction formalism.
Although the discussion in Ref.~\cite{Melnikov:2011qx} was restricted to the case of the top-quark decay, the subtraction terms given there should in principle be applicable to the decay of any strongly interacting particle, however, with the restriction that only massless partons appear in the final state.

So far, no general algorithm exists for the calculation of NLO QED corrections to decay processes using the dipole subtraction formalism.
In this paper we fill this gap by explicitly constructing such a process-independent subtraction term considering both massless and massive final-state fermions 
and work out its extension to the treatment of non-collinear-safe observables.
The further extension of the formalism to treat massive partons in NLO QCD corrections to decay processes is then fully straightforward.

The \emph{one-cutoff phase-space slicing} (OCS) method,
considered in the second part of the paper, 
was initially introduced in Ref.~\cite{Giele:1991vf} for the calculation of NLO QCD corrections to the multi-jet production cross section in $\Pep\Pem$ annihilation.
As such, its application was limited to the case with only massless coloured partons in the final state. It was later extended to allow for incoming partons~\cite{Giele:1993dj} and massive quarks~\cite{Keller:1998tf}.
The OCS method is distinguished by the feature that one single Lorentz-invariant cut is introduced to isolate the soft and/or collinear regions in the real-emission phase space.
Another frequently used slicing method, which is not considered in detail in this paper, is the so-called two-cutoff slicing method~\cite{Harris:2001sx,Harris:2002md}.
As a consequence of introducing two independent cuts on energies and angles, however, this approach is not manifestly Lorentz invariant.

In its original formulation, the OCS method is restricted to amplitudes that involve only two sources (particles) of IR singularities, a feature that is naturally supported by colour-ordered QCD amplitudes, but neither by sub-leading colour structures in QCD nor 
by IR singularities originating from $U(1)$ gauge bosons, such as the photon in QED.
Furthermore, no results employing mass regularization are documented
in the literature, although it is a common choice in the calculation of electroweak radiative corrections.
In this paper we extend the 
OCS formalism in these two respects, i.e.\ presenting a formulation employing mass regularization 
and dealing with an arbitrary number of soft or collinear singularities.
To this end, we employ results from the aforementioned dipole subtraction formalism to obtain an approximation of the real-emission matrix elements in the singular regions together with the associated phase-space parametrizations suitable for the subsequent analytic integration.
Furthermore, we discuss the subtleties that arise in the case of processes that involve arbitrary many sources of IR singularities.
\vspace{.5em}

At its design luminosity in Run~2, the LHC will be a top-quark factory, copiously producing top-quarks in pairs, as single particles, or in association with other particles.
As the heaviest elementary particle known to date, the top-quark is a unique window to explore the mechanism of electroweak symmetry breaking and to test the Standard Model (SM) on the one hand, and to probe physics beyond the SM on the other. 
Apart from the top-quark mass, key observables in the top sector comprise the top-quark width and all accessible production cross sections, for which precise theoretical predictions are required.
In these predictions the top-quark width $\Gamma_{\Pqt}$ 
plays a double role: 
as an observable in its own right and as an ingredient in cross-section predictions that take into account top-quark decays.
In the latter case, $\Gamma_{\Pqt}$ enters the Breit--Wigner propagator of the top-quark resonance, so that it should be known to the same perturbative order as the calculated cross section.

The corrections to the top-quark decay process $\Pqt\to\Pqb\PWp$ are known at 
NLO~\cite{Jezabek:1988iv, Ghinculov:2000nx} and next-to-next-to-leading order (NNLO)~\cite{Czarnecki:1995jt,Czarnecki:1998qc, Chetyrkin:1999ju,Gao:2012ja} in QCD and up to NLO for the electroweak (EW) corrections~\cite{Denner:1990ns, Denner:1991kt, Eilam:1991iz, Jezabek:1993wk, Oliveira:2001vw}.
However, all prior calculations either consider on-shell $\PW$ bosons or only partially account for off-shell effects by performing an expansion  
about the intermediate W~resonance.
As an application of the techniques introduced in this work, we present the fully differential calculation of the QCD and EW corrections at NLO to the three-body decay of the top-quark, accounting for off-shell effects of the intermediate $\PW$ boson within the 
complex-mass scheme~\cite{Denner:1999gp, Denner:2005fg, Denner:2006ic} and retaining 
a finite bottom-quark mass. 
We consider both the hadronic and semi-leptonic decay channels and further allow for non-collinear-safe photon emission (``bare leptons'') for the latter.
The presented calculation of the NLO EW corrections to the decays $\Pqt\to\Pqb\PWp\to\Pqb\Plp\Pgnl / \Pqb\Paq\Pq'$, thus, provides an ingredient to the (not yet existing) calculation of NLO EW corrections to the $\Pqt\Paqt$ production process 
$\Pp\Pp\to\Pqb\Paqb\PWp\PWm\to6\Pf$ including top-quark decays, similar to the respective NLO QCD corrections to the top-quark decay in the NLO QCD calculation to $\Pp\Pp\to\Pqb\Paqb\PWp\PWm\to6\Pf$~\cite{Denner:2010jp,Bevilacqua:2010qb,Denner:2012yc,Cascioli:2013wga}.

On the experimental side, the first direct experimental determination of the top-quark decay width was performed by the experiments at the Tevatron collider~\cite{Aaltonen:2013kna,Abazov:2012vd}. 
It, however, still suffered from large experimental uncertainties.
Recently, the first measurement 
at the LHC was presented by the CMS experiment with the data collected at $\sqrt{s}=8~\TeV$~\cite{Khachatryan:2014nda},  significantly improving on the previous measurements, with a precision at the $10\%$ level.

This paper is organized as follows:
In Sect.~\ref{sec:nlo} we briefly review the general structure of an NLO calculation to decay processes to set up the notation and conventions used throughout this work.
Section~\ref{sec:dipole} first outlines the dipole subtraction formalism in Sect.~\ref{sec:dipole:overview} and subsequently discusses its extension to deal with decay kinematics.
Massive final-state fermions are treated in Sect.~\ref{sec:dipole:massive}, and the special case of light fermions is covered in Sect.~\ref{sec:dipole:light+kabelschacht}, where we also discuss the modifications for non-collinear-safe observables.
The OCS method is presented in Sect.~\ref{sec:ocsm} with a general overview of the formalism in Sect.~\ref{sect:OCSMoverview}, 
explaining in particular the origin and treatment of so-called ``surface terms'' which arise in the case of processes with more than two sources of IR singularities.
The various cases that arise from the different combinations of massless or massive particles in the initial or final states are covered in Sect.~\ref{sect:OCSM_IR}. 
The application of the two techniques to the top-quark decay at NLO is presented in Sect.~\ref{sec:top} where 
the calculational details are summarized in Sect.~\ref{sec:top:setup}.
In Sect.~\ref{sec:top:total} and \ref{sec:top:diff} we present the numerical results for the corrections to the decay width and the differential distributions, respectively.
The conclusions and an outlook are given in Sect.~\ref{sec:concl}.


\section{NLO corrections to decay processes---notation and conventions}
\label{sec:nlo}

The leading-order (LO) partial decay width of a particle with mass $m_a$ decaying into a specific final-state configuration comprising $n$ particles is given by 
\begin{align}
	\label{eq:gamma-0}
	\Gamma^{(0)} 
	&=
	\frac{1}{2p_a^0}\int\rd\lips{n}\, \frac{1}{S_{\{n\}}} 
	\avgsum \lvert \cM_0(\lips{n}) \rvert^2  ,
\end{align}
where $\cM_0$ is the lowest-order transition amplitude for the $1 \to n$ decay process with $\avgsum$ representing the possible sum (average) over the final (initial) state degrees of freedom and $S_{\{n\}}$ is the symmetry factor for identical particles in the final state.
Here we have introduced the abbreviation 
\begin{equation}
	\label{eq:lips}
	\rd\lips{n} \equiv
	\rd\phi(p_a;p_1,\ldots,p_n) 
	=\left(\prod_{i=1}^n\frac{\rd^4 p_i}{(2\pi)^3} \; \theta(p_i^0) \; \delta(p_i^2-m_i^2)\right)
	(2\pi)^4 \delta\left(\sum_{i=1}^n p_i - p_a\right)
\end{equation}
for the $n$-particle phase-space measure,
where $p_i$ and $m_i$ denote the momentum and mass of the $i$-th final-state particle, respectively, and $p_a$ the momentum of the decaying particle, with $p_a^\mu=(m_a,\vec{0})$ in its centre-of-mass frame.
Furthermore, we have used the shorthand notation 
$\lips{n}\equiv\{p_1,\ldots,p_n\}$
for the associated kinematics given by the set of momenta of the particles. 

At NLO, the decay width additionally receives contributions from radiative corrections,
\begin{align}
	\label{eq:gamma-nlo}
	\Gamma^\NLO = \Gamma^{(0)} + \Gamma^{(1)} ,
\end{align}
with
\begin{align}
	\label{eq:gamma-1}
	\Gamma^{(1)} &=
	\frac{1}{2p_a^0}\int\rd\lips{n} \frac{1}{S_{\{n\}}} 
	\avgsum 2\Re\Big\{ \delta\cM(\lips{n}) \big(\cM_0(\lips{n})\big)^* \Big\}
	\nonumber\\&\quad
	+\frac{1}{2p_a^0}\int\rd\lips{n+1} \frac{1}{S_{\{n+1\}}} 
	\avgsum \lvert \cM_1(\lips{n+1}) \rvert^2 ,
\end{align}
where $\delta\cM$ and $\cM_1$ denote the virtual and real-emission matrix elements, respectively.
Here we assume that the virtual corrections already include the counterterm contributions, so that all ultraviolet singularities are properly cancelled after renormalization.

The calculation of higher-order corrections involving the emission of massless particles, in general, leads to the occurrence of IR singularities in both the virtual and real corrections which cancel in the sum of Eq.~\eqref{eq:gamma-1} by virtue of the 
KLN theorem.%
\footnote{In case of processes with identified hadrons in the initial and/or final state, 
collinear singularities arise for which the inclusiveness assumptions of the KLN theorem 
are not fulfilled. These universal singularities can be absorbed into a redefinition of the corresponding (NLO) structure functions given by the parton distribution functions and the fragmentation functions, respectively.}
Although the cancellation of IR singularities is well understood, 
in practise, the procedure
is non-trivial, since the virtual and real corrections are defined on different phase spaces.
Moreover, the divergences contained in the real corrections are implicitly hidden in the phase-space integration, whereas the singularities in the virtual corrections are made explicit using regulators when performing the integration over the loop momentum.
To accomplish an analytic cancellation of all IR singularities therefore requires to make the singularities of the real corrections explicit by means of regulators.
On the other hand, it is desirable to retain the flexibility of a fully differential numerical approach for the phase-space integration.
Both the slicing and subtraction techniques provide a prescription to perform a fully differential calculation suitable for a numerical evaluation while having the full analytic control over the IR singularities and their cancellation.
Both approaches exploit the factorization properties of the real-emission amplitude in the various singular limits which we briefly review in the following for the case of photon radiation off fermions in order to set up our notation.
For further details we refer to Ref.~\cite{Dittmaier:1999mb} which we follow closely.
Throughout this work we regularize the IR singularities using \emph{mass regularization} where an infinitesimal photon mass $m_{\Pgg}$ is introduced and the masses of the light fermions $m_{\Pf}$ are retained in the divergent logarithms.
A prescription for the conversion of the results obtained using mass regularization to the corresponding expressions using dimensional regularization is given in the Appendix.

The singular behaviour of emission amplitudes 
in the \emph{soft-photon limit} ($k\to0$) can be described by the well-known eikonal approximation where the squared real-emission amplitude, summed over the photon polarizations $\lambda_{\Pgg}$, can be written as follows,
\begin{align}
	\label{eq:ir-soft}
	\sum_{\lambda_{\Pgg}} \lvert \cM_1(k) \rvert^2
	\;&\widesim[3]{k\to0}\;
	- \sum_{\Pf,\Pf'} \sigma_{\Pf}Q_{\Pf} \, \sigma_{\Pf'}Q_{\Pf'} \, e^2
	\,\frac{ p_{\Pf}p_{\Pf'} }{(p_{\Pf}k)(p_{\Pf'}k)}
	\,\lvert \cM_0 \rvert^2 ,
\end{align}
where $e$ is the elementary charge.
Here the summation over $\Pf,\Pf'$ extends over all charged external particles (of any spin), 
and the reduced matrix element $\cM_0$ is evaluated with the set of momenta that is obtained by omitting the momentum $k$ of the soft photon.
We have further introduced the sign factors $\sigma_{\Pf}=\pm1$ describing the charge flow, which are defined as $\sigma_{\Pf}=+1$ for incoming particles and outgoing anti-particles and $\sigma_{\Pf}=-1$ for incoming anti-particles and outgoing particles.
Overall charge conservation can then be expressed as
\begin{equation}
	\sum_{\Pf} \sigma_{\Pf} Q_{\Pf} = 0 .
\end{equation}

For the case of decay processes considered here, the mass $m_a$ of the decaying particle represents the maximal scale in 
the process and therefore will never be considered in the small-mass limit.
As a consequence, only final-state collinear singularities arise where the radiated photon becomes collinear to a light fermion in the final state.
The asymptotic behaviour of the squared transition matrix in this limit takes the form
\begin{align}
	\label{eq:ir-collinear}
	\sum_{\lambda_{\Pgg}} \lvert \cM_1(p_i,k;\kappa_i) \rvert^2
	\;&\widesim[3]{p_ik\to0}\; 
	Q_i^2\,e^2\,
	g_{i,\tau}^\text{(out)}(p_i,k)
	\,\lvert \cM_0(p_i+k,\tau\kappa_i) \rvert^2 , 
\end{align}
where $\kappa_i$ denotes the helicity of particle $i$ which might change to any other
spin value $\tau\kappa_i$. If $i$ is a spin-$\frac{1}{2}$ fermion, the spin can only
potentially change sign, i.e.\ there is just a sum over $\tau=\pm$. In this case,
the functions $g_{i,\tau}^\text{(out)}$ are given by
\begin{subequations}
\begin{align}
	g_{i,+}^\text{(out)}(p_i,k) &=
	\frac{1}{p_ik}
	\left[\frac{1+z_i^2}{1-z_i}-\frac{m_i^2}{p_ik}\right]
	-g_{i,-}^\text{(out)}(p_i,k) ,
	\\
	g_{i,-}^\text{(out)}(p_i,k) &=
	\frac{m_i^2}{2(p_ik)^2} \, \frac{(1-z_i)^2}{z_i} ,
\end{align}
\label{eq:gout}
\end{subequations}
with the dimensionless variable
\begin{align}
	z_i &= \frac{p_i^0}{p_i^0+k^0}
	\label{eq:z_i}
\end{align}
representing the momentum fraction carried away by the fermion after the collinear splitting $\Pf\to\Pf+\Pgg$.
For the case of unpolarized fermions, Eq.~\eqref{eq:ir-collinear} reduces to
\begin{align}
	\label{eq:ir-collinear-unpol}
	\sum_{\lambda_{\Pgg}} \lvert \cM_1(p_i,k) \rvert^2
	\;&\widesim[3]{p_ik\to0}\; 
	Q_i^2\,e^2\,
	g_{i}^\text{(out)}(p_i,k)
	\,\lvert \cM_0(p_i+k) \rvert^2 ,
\end{align}
with $g_{i}^\text{(out)}\equiv g_{i,+}^\text{(out)} + g_{i,-}^\text{(out)}$.\\

Note that Eq.~\eqref{eq:gout} does not describe
particles with a spin other than $\frac{1}{2}$. However, any subtraction formalism 
based on this asymptotics may still be used for particles with another spin 
(such as a W~boson) as long
those particles are heavy enough to avoid collinear singularities, because the 
asymptotic behaviour \eqref{eq:ir-soft} in the soft-photon limit is spin independent.


\section{The dipole subtraction formalism}
\label{sec:dipole}

As was pointed out in the introduction, so far, no general formalism exists yet in the literature for the calculation of NLO QED corrections to decay processes within the dipole subtraction formalism.
We present the construction of such a general subtraction term in the following and begin in Sect.~\ref{sec:dipole:overview} with a brief overview of the formalism and of our approach for its generalization to decay processes.
The explicit form of the corresponding radiator function and its integrated counterpart is given in Sect.~\ref{sec:dipole:massive} for the case of a massive final-state fermion.
Section~\ref{sec:dipole:light+kabelschacht} considers the case of light fermions and the extension to non-collinear-safe observables.

The described formalism was already successfully applied in the evaluation of the
NLO EW and mixed NNLO QCD/EW corrections to Drell--Yan-like W- and Z-boson production
in the resonance region presented in Refs.~\cite{Dittmaier:2014qza,Dittmaier:2014koa},
where the corrections to the gauge-boson decay processes appear as building blocks
of the full calculation.

\subsection{Overview of the method}
\label{sec:dipole:overview}

The general idea of the subtraction formalism can be summarized in the following formula,
\begin{equation}
	\label{eq:subtraction-general}
	\int\rd\lips{n+1}\sum_{\lambda_{\Pgg}}\modsq{\cM_1} = 
	\int\rd\lips{n+1} 
	\Biggl( \sum_{\lambda_{\Pgg}}\modsq{\cM_1} - \modsq{\cMsub} \Biggr)
	+ \int\rd\lips[]{n} \otimes \left(\int[\rd k] \modsq{\cMsub}\right) ,
\end{equation}
where $\cM_1$ denotes the transition amplitude of the bremsstrahlung process.
The auxiliary term $\modsq{\cMsub}$ is specifically tailored to act as a local counterterm to the real-emission corrections.
The subtraction term \modsq{\cMsub} point-wise mimics the real-emission contributions in all its singular limits, 
i.e.\ it satisfies
the following properties,
\begin{align}
	\label{eq:subtraction-general-limits}
	\modsq{\cMsub} &\widesim{} \sum_{\lambda_{\Pgg}}\modsq{\cM_1}  &
	\text{for\quad $k\to0$ \quad or\quad ${p_i}{k}\to0$} ,
\end{align}
where $p_i$ denotes the four-momentum associated with a light fermion in the final state and $\sum_{\lambda_{\Pgg}}$ indicates the summation over the photon polarization.
The explicit asymptotics in the respective limits are given in Eqs.~\eqref{eq:ir-soft} and \eqref{eq:ir-collinear}.
The property~\eqref{eq:subtraction-general-limits} ensures that the first term on the r.h.s.\ of Eq.~\eqref{eq:subtraction-general} is free of IR singularities.
As a result, the integration over the full real-emission phase space $\rd\lips{n+1}$ can be performed without regulators which makes it suitable for the
evaluation using numerical methods.
The singularities originally contained in the real-emission corrections are now fully encapsulated in the second term of Eq.~\eqref{eq:subtraction-general} which is constructed in such a way to allow for a direct analytical integration of the singular contributions. 
To this end, the real-emission phase space is factorized into the singular sub-space given by the photonic part $[\rd k]$ and the remaining phase space integration $\rd\lips[]{n}$ associated with the non-radiative process.  
The symbol ``$\otimes$'' indicates that this phase-space factorization may contain additional convolutions or summations.
Performing the analytical integration over $[\rd k]$ 
results in an expression where the singularities are explicitly regularized in some appropriate regularization scheme. 
These divergences can then be analytically combined with
the respective singularities appearing in the virtual corrections, so that possible cancellations
can be made explicit.
The remaining integration $\rd\lips[]{n}$ is free of singularities and can be evaluated numerically.

The dipole subtraction formalism exploits the factorization 
of $\sum_{\lambda_{\Pgg}}\modsq{\cM_1}$ in the singular limits
by partitioning the subtraction term $\modsq{\cMsub}$ into so-called \emph{dipoles} $\modsq{\cMsub[\Pf\Pf']}$ according to
\begin{align}
	\label{eq:dipole-subtraction-general}
	\modsq{\cMsub(\lips{n+1})} &= 
	\sum_{\substack{\Pf,\Pf'\\\Pf\ne \Pf'}} \modsq{\cMsub[\Pf\Pf'](\lips{n+1})}
	\nonumber\\&=
	-e^2 \sum_{\substack{\Pf,\Pf'\\\Pf\ne \Pf'}} \sigma_\Pf Q_\Pf\, \sigma_{\Pf'} Q_{\Pf'}\;
	\gsub{\Pf\Pf',\tau}(p_\Pf,p_{\Pf'},k)\,
	\modsq{\cM_0(\lips[\Pf\Pf']{n};\tau\kappa_{\Pf})} ,
\end{align}
where $\Pf$ and $\Pf'$ are called the \emph{emitter} and \emph{spectator} particles, respectively.
The dipole terms are
written as a product of the auxiliary functions $\gsub{\Pf\Pf'}$ and the squared matrix element $\modsq{\cM_0}$, where the former represent universal functions that encapsulate the singularity structure and can be integrated analytically over the singular photon phase space once and for all. 
The associated squared matrix element $\modsq{\cM_0}$ is evaluated on the reduced $n$-particle phase space \lips[\Pf\Pf']{n} that remains after isolating the singular one-particle sub-space $[\rd k]$ of the photon.
These \emph{dipole mappings} are required to reproduce the appropriate kinematic configuration in the singular regions and further respect exact momentum conservation as well as all mass-shell conditions.

\begin{figure}
	\centering
	\includegraphics[scale=1.25]{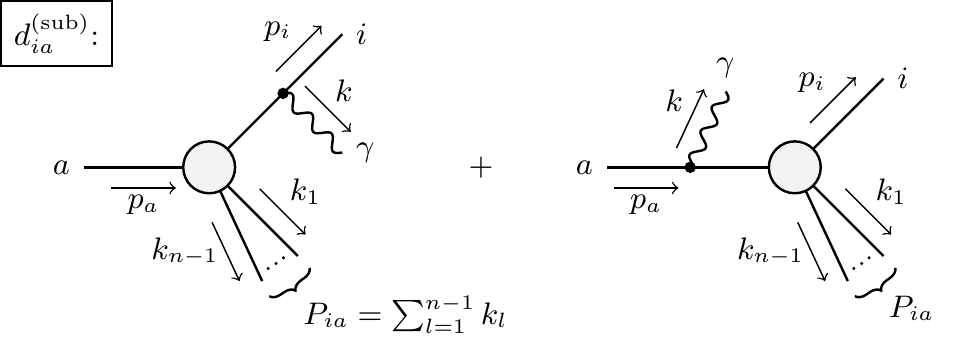}
	\caption{Schematic diagrams illustrating the subtraction function \dsub{ia} for the decay of a particle $a$ and a charged final-state particle $i$. 
	The decay dipole merges the two configurations associated with photon radiation from the final~(left) and initial~(right) state into one subtraction term.}
	\label{fig:dipole-qed-decay}
\end{figure}

A plain extension in the spirit of the dipole subtraction formalism for the decay process would construct two new subtraction terms: 
a dipole where the decaying particle $a$ is 
the emitter, \gsub{ai}, and second where $a$ acts as the spectator \gsub{ia}.
However, since there are no collinear singularities associated with the initial-state particle $a$,
we combine the two subtraction terms to a single term called 
$\dsub{ia}= \gsub{ia}+ \gsub{ai}$
which is illustrated in Fig.~\ref{fig:dipole-qed-decay} in terms of the two schematic diagrams that would be associated with \gsub{ia} and \gsub{ai}.

Such a combined treatment reduces the number of independent dipoles that need to be computed and, moreover, allows part of the calculation to be taken over from the 
scattering case with a final-state emitter and spectator described in Ref.~\cite{Dittmaier:1999mb}.
The asymptotic behaviour of the subtraction function \dsub{ia} differs from those of 
\gsub{ia} and \gsub{ai}
of Ref.~\cite{Dittmaier:1999mb} in the soft limit and explicitly reads
\begin{equation}
\begin{aligned}
	\dsub{ia,+}(p_i,p_a,k) &\;\widesim[3]{k\to0}\;
	\frac{2{p_i}{p_a}}{({p_i}{k})({p_a}{k})}
	-\frac{m_i^2}{({p_i}{k})^2} -\frac{m_a^2}{({p_a}{k})^2} ,
	\\
	\dsub{ia,-}(p_i,p_a,k) &\;\widesim[3]{k\to0}\;
	\order{1}.
\end{aligned}
\label{eq:dipole-decay-soft-limit}
\end{equation}
In the limit of small fermion masses $m_i\to0$ the asymptotic form in the collinear region is required to be
\begin{align}
	\label{eq:dipole-decay-collinear-limit-massreg}
	\dsub{ia,\tau}(p_i,p_a,k) &\;\widesim[3]{{p_i}{k}\to0}\;
	g_{i,\tau}^\text{(out)}(p_i,k)
\end{align}
with the functions $g_{i,\pm}^\text{(out)}$ given in Eq.~\eqref{eq:gout}.

The subtraction term \modsq{\cMsub} in the dipole subtraction formalism is then given by the following expression in the case of decay processes, cf.\ Eq.~\eqref{eq:dipole-subtraction-general},
\begin{align}
	\label{eq:dipole-subtraction-general:decay}
	\modsq{\cMsub(\lips{n+1})} &= 
	\sum_i \modsq{\cMsub[ ia ](\lips{n+1})} 
	+\sum_{\substack{i,j\\i\ne j}} \modsq{\cMsub[ij](\lips{n+1})}
	\nonumber\\&=
	-e^2 \sum_i \sigma_i Q_i \,
	\Biggl\{ \sigma_a Q_a\,\dsub{ia,\tau}(p_i,p_a,k) \modsq{\cM_0(\lips[ia]{n};\tau\kappa_i)}
	\nonumber\\&\qquad
	+\sum_{\mathclap{\substack{j \\ j\ne i}}} \sigma_j Q_j\, \gsub{ij,\tau}(p_i,p_j,k)
	\modsq{\cM_0(\lips[ij]{n};\tau\kappa_i)} \Biggr\} ,
\end{align}
where the summation over the indices $i,j$ extends over all charged particles in the 
final state, $a$ denotes the decaying particle, and \gsub{ij} are the known subtraction functions of the dipoles with a final-state emitter--spectator pair given in Ref.~\cite{Dittmaier:1999mb}.

\subsection{Subtraction function for massive final-state fermions}
\label{sec:dipole:massive}

In the following we present the construction of the subtraction function \dsub{ia} for the decay of a particle $a$ of mass $m_a$ with a massive final-state fermion $i$ of mass $m_i$. 
The special case of light fermions can be obtained by a systematic expansion in $m_i$ and will be discussed in the next section together with the extension to non-collinear-safe observables.

In the construction of the phase-space mapping \lips[ia]{n} for the decay dipole it is convenient to keep the four-momentum of the decaying particle $a$ fixed.
This constraint can be maintained by the deformation of the momenta $k_l$ of all particles that are not involved in the respective dipole term via a Lorentz transformation.
Such a transformation is necessary in order to compensate the recoil induced by the radiated photon so that four-momentum conservation and all mass-shell conditions are retained.
To this end, we define the auxiliary momentum $P_{ia}$ of the ``recoiling system'', see Fig.~\ref{fig:dipole-qed-decay},
\begin{equation}
	P_{ia}^\mu = p_a^\mu -p_i^\mu -k^\mu = \sum_{l=1}^{n-1} k_l^\mu ,
\end{equation}
which absorbs the recoil of the radiated photon through the transformation
\begin{align}
	\widetilde{P}_{ia}^\mu &=
	\frac{\sqrt{\lambda_{ia}}}{\sqrt{\lambda((p_i+k)^2,m_a^2,P_{ia}^2)}}
	\left( P_{ia}^\mu -\frac{{P_{ia}}{p_a}}{m_a^2}\, p_a^\mu \right)
	+\frac{m_a^2+P_{ia}^2-m_i^2}{2 m_a^2} p_a^\mu ,
\end{align}
where 
\begin{align}
	\lambda_{ia} &= \lambda(m_i^2,m_a^2,P_{ia}^2) , 
\end{align}
and $\lambda(x,y,z)$ denotes the K\"all\'en function defined as
\begin{equation}
 \lambda(x,y,z)=x^2+y^2+z^2-2xy-2xz-2yz .
 \label{eq:kallen}
\end{equation}
We further introduce the quantity
\begin{equation}
	\bar{P}_{ia}^2 = m_a^2 - m_i^2 - P_{ia}^2 -m_{\Pgg}^2,
\end{equation}
where $m_{\Pgg}$ denotes the infinitesimal photon mass that is introduced as a regulator for the soft singularities for later convenience.
The phase-space mapping of the decay dipole then explicitly reads
\begin{align}
	\label{eq:dipole-decay-mapping}
	\tilde{p}_a^\mu &= p_a^\mu , & 
	\tilde{p}_i^\mu &= p_a^\mu - \widetilde{P}_{ia}^\mu , &
	\tilde{k}_l^\mu &= \Lambda^\mu{}_\nu\, k_l^\nu ,
\end{align}
where $k_l$ denotes the momenta of all remaining particles of the decay process, including all neutral particles.
The Lorentz transformation $\Lambda^\mu{}_\nu$ can be expressed in terms of the momenta $P^\mu_{ia}$ and $\widetilde{P}^\mu_{ia}$ as follows,
\begin{align}
	\Lambda^\mu{}_\nu &= g^\mu{}_\nu 
	-\frac{(P_{ia}+\widetilde{P}_{ia})^\mu\;(P_{ia}+\widetilde{P}_{ia})_\nu}
	{P_{ia}^2+{P_{ia}}{\widetilde{P}_{ia}}}
	+\frac{2\widetilde{P}_{ia}^\mu \; P_{ia,\nu}}{P_{ia}^2} .
\end{align}
This phase-space parametrization is identical to the one used in the 
scattering case with a final-state emitter 
and a final-state spectator $j$ 
of Ref.~\cite{Dittmaier:1999mb}, where particle $i$ plays the role of the emitter and 
the recoiling system the role of the spectator of mass $m_j^2=P_{ia}^2=\widetilde{P}_{ia}^2$.
In detail, the following substitutions have to be made in order to transfer the
quantities of Section~4.1 of Ref.~\cite{Dittmaier:1999mb} to the kinematics considered here,
\begin{align}
p_j\to P_{ia}, \quad
m_j^2\to P_{ia}^2,\quad
\tilde p_j \to \tilde P_{ia},\quad
P_{ij}\to p_a,\quad
P_{ij}^2\to m_a^2,\quad
\bar P_{ij}^2\to \bar{P}_{ia}^2.
\end{align}
Introducing the dipole variables 
\begin{align}
	\label{eq:dipole-decay-variables}
	y_{ia} &=
	\frac{ 2{p_i}{k} }{ \bar{P}_{ia}^2 } , &
	z_{ia} &=
	\frac{ {p_i}{P_{ia}} }{ {p_i}{P_{ia}} +{P_{ia}}{k} } ,
\end{align}
we construct the subtraction functions \dsub{ia,\tau} as follows,
\begin{align}
	\dsub{ia,+}(p_i,p_a,k) 
	&= \frac{1}{({p_i}{k}) \, R_{ia}(y_{ia})} 
	\left[ \frac{2}{1-z_{ia}(1-y_{ia})} \left(1+\frac{2 m_i^2}{\bar{P}_{ia}^2}\right)
	-1-z_{ia} -\frac{m_i^2}{{p_i}{k}} \right] 
	\nonumber\\*&\quad
	-\frac{m_a^2}{\bar{P}_{ia}^4 \, R_{ia}(y_{ia})}
	\,\frac{4}{[1-z_{ia}(1-y_{ia})]^2} 
	- \dsub{ia,-}(p_i,p_a,k),
	\nonumber\\
	\dsub{ia,-}(p_i,p_a,k) 
	&= \frac{m_i^2}{2(p_ik)^2} \, \frac{(1-z_{ia})^2}{z_{ia}} \, \frac{r_{ia}(y_{ia})}{R_{ia}(y_{ia})}	,
\label{eq:dipole-decay-dia}
\end{align}
where we have defined the auxiliary functions
\begin{align}
	R_{ia}(y_{ia}) &= 
	\frac{ \sqrt{\bigl[ 2 P_{ia}^2 +\bar{P}_{ia}^2(1-y_{ia}) \bigr]^2-4 m_a^2P_{ia}^2} }
	{ \sqrt{\lambda_{ia}} } , 
	\\
	r_{ia}(y_{ia}) &= 1-\frac{2P_{ia}^2 \bigl(2 m_i^2+\bar{P}_{ia}^2 \bigr)}{\lambda_{ia}}\,\frac{y_{ia}}{1-y_{ia}} .
\end{align}
Note the similarity of these subtraction terms to the final--final dipoles \gsub{ij} of Ref.~\cite{Dittmaier:1999mb} up to modifications that establish the appropriate behaviour in the soft limit~\eqref{eq:dipole-decay-soft-limit}.
In detail, the functions $g_{ij,\tau}$ defined in Eq.~(4.4) of Ref.~\cite{Dittmaier:1999mb} coincide with
$\dsub{ia,\tau}$ up to the first term in the second line of Eq.~\eqref{eq:dipole-decay-dia},
which is new, and by an additional factor of $(1+2m_i^2/\bar{P}_{ia}^2)$ in the eikonal term
(first term in the first line).
As a consequence, many parts of the calculations can be taken over from Ref.~\cite{Dittmaier:1999mb}, considerably simplifying the analytic integration over the photonic part of the phase space.
It is easy to verify that the radiator function \dsub{ia} defined in Eq.~\eqref{eq:dipole-decay-dia} reproduces the correct asymptotic behaviour in the soft and collinear limits given in Eqs.~\eqref{eq:dipole-decay-soft-limit} and \eqref{eq:dipole-decay-collinear-limit-massreg}, respectively.

The analytic integration of the subtraction functions over the photonic part of the phase space gives rise to the integrated counterparts 
\begin{align}
	\Dsub{ia,\tau}(\lips[]{n}) &=
	8 \pi^2 \int [\rd k(m_a^2,y_{ia},z_{ia})]\; \dsub{ia,\tau}(p_i,p_a,k) .
\end{align}
The integrated decay dipole can then be written as
\begin{align}
	\label{eq:dipole-dipole:integral}
	\int\rd\lips{n+1}\modsq{\cMsub[ia](\lips{n+1})}
	= -\frac{\alpha}{2\pi}\,\sigma_iQ_i\,\sigma_aQ_a 
	\int\rd\lips[]{n}\; \Dsub{ia,\tau}(\lips[]{n}) \modsq{\cM_0(\lips[]{n};\tau\kappa_i)} ,
\end{align}
with $\alpha=e^2/(4\pi)$ and
\begin{align}
	\Dsub{ia,+}(\lips[]{n}) &=
	 2\ln\left(\frac{m_{\Pgg}^2}{\bar{P}_{ia}^2}\right)
	+3\ln(a_3) -2\ln(1-a_3^2)
	-2\ln\left(\frac{(m_a-\sqrt{P_{ia}^2})^2-m_i^2}{\bar{P}_{ia}^2}\right)
	+\frac{a_3^2}{2} +\frac{3}{2}
	\nonumber\\*&\quad
	+\frac{\bar{P}_{ia}^2+2 m_i^2}{\sqrt{\lambda_{ia}}} \Biggl\{
	\ln(a_1)\ln\left(\frac{m_{\Pgg}^2 P_{ia}^2}{\lambda_{ia} a_2}\right)
	+2\Li_2(a_1) +4\Li_2\left(-\sqrt{\frac{a_2}{a_1}}\right)
	\nonumber\\*&\qquad
	-4\Li_2\left(-\sqrt{a_1 a_2}\right)
	+\frac{1}{2}\ln^2(a_1)
	-\ln(a_1)
	-\frac{\pi^2}{3} 
	\Biggr\} 
	-\Dsub{ia,-}(\lips[]{n}),
	\nonumber\\
	\Dsub{ia,-}(\lips[]{n}) &=
	\frac{\bar{P}_{ia}^4}{\lambda_{ia}} \Biggl\{
	\frac{2 m_i^2}{\bar{P}_{ia}^2}\ln(a_3) 
	+\frac{4 m_i^2 P_{ia}^2}{\bar{P}_{ia}^4}\ln\left(\frac{1+a_3^2}{2 a_3}\right)
	+\frac{ m_i^2 P_{ia}^2}{\bigl(\bar{P}_{ia}^2+ m_i^2\bigr)^2}\ln\left(\frac{2 m_i \sqrt{P_{ia}^2}a_3}{\bar{P}_{ia}^2}\right)
	\nonumber\\*&\qquad
	+\frac{ m_i \sqrt{P_{ia}^2}\bigl(\bar{P}_{ia}^2+2 m_i^2\bigr)}{\bar{P}_{ia}^4}[4\arctan(a_3)-\pi]
	\nonumber\\*&\qquad
	+(1-a_3^2)\left[\frac{1}{2}+\frac{ m_i^2 m_a^2}{\bar{P}_{ia}^2(\bar{P}_{ia}^2+ m_i^2)}+\frac{2 m_i^3 \sqrt{P_{ia}^2}}{a_3\bar{P}_{ia}^4}\right]
	\Biggr\} ,
\label{eq:dipole-decay-Dia}
\end{align}
and the abbreviations
\begin{align}
	\label{eq:integrated-dipole-decay-abbr}
	a_1 &= 
	\frac{\bar{P}_{ia}^2+2 m_i^2-\sqrt{\lambda_{ia}}}{\bar{P}_{ia}^2+2 m_i^2+\sqrt{\lambda_{ia}}} , &
	a_2 &=
	\frac{\bar{P}_{ia}^2-\sqrt{\lambda_{ia}}}{\bar{P}_{ia}^2+\sqrt{\lambda_{ia}}} , &
	a_3 &= 
	\frac{ m_i}{m_a-\sqrt{P_{ia}^2} } .
\end{align}

\subsection{Light fermions and non-collinear-safe observables}
\label{sec:dipole:light+kabelschacht}

The subtraction functions \dsub{ia,\tau} given in Eq.~\eqref{eq:dipole-decay-dia} 
are already defined in such a way to correctly reproduce the asymptotic form in the collinear limit as demanded in Eq.~\eqref{eq:dipole-decay-collinear-limit-massreg} for the special of case of small fermion masses $m_i\to0$.
The corresponding expressions for the integrated dipoles 
are obtained by expanding Eq.~\eqref{eq:dipole-decay-Dia} for small $m_i$ and 
omitting mass-suppressed terms, 
\begin{align}
	\Dsub{ia,+}(\lips[]{n}) &=
        {\cal L}(\bar{P}_{ia}^2,m_i^2)
        +{\cal L}(\bar{P}_{ia}^2,m_a^2)
	+4\Li_2\left(-\frac{\sqrt{P_{ia}^2}}{m_a}\right) 
	+\ln^2\left(\frac{\bar{P}_{ia}^2}{m_a^2}\right)
	\nonumber\\*&\quad
	+2\ln\left(\frac{m_a+\sqrt{P_{ia}^2}}{m_a-\sqrt{P_{ia}^2}}\right)
	+3\ln\left(\frac{m_a+\sqrt{P_{ia}^2}}{m_a}\right) 
	-\frac{\pi^2}{3} +\frac{3}{2} 
	-\Dsub{ia,-}(\lips[]{n}) ,
	\nonumber\\
	\Dsub{ia,-}(\lips[]{n}) &=
	\frac{1}{2} ,
\label{eq:dipole-decay-Dia:massless-massreg}
\end{align}
where we made use of the auxliliary function
\begin{align}
{\cal L}(P^2,m^2) =
\ln\biggl(\frac{m^2}{P^2}\biggr)
\ln\biggl(\frac{m_\gamma^2}{P^2}\biggr)
+ \ln\biggl(\frac{m_\gamma^2}{P^2}\biggr)
- \frac{1}{2}\ln^2\biggl(\frac{m^2}{P^2}\biggr)
+ \frac{1}{2}\ln\biggl(\frac{m^2}{P^2}\biggr),
\label{eq:L}
\end{align}
which was introduced in Eq.~(3.8) of Ref.~\cite{Dittmaier:1999mb} to 
describe the universal singular behaviour of the integrated dipole functions
\Gsub{}.

The corresponding result where the IR singularities are regularized using dimensional regularization can be obtained from Eq.~\eqref{eq:dipole-decay-Dia:massless-massreg} by performing a transition between the regularization schemes using the results of Ref.~\cite{Catani:2000ef} which is briefly summarized in the Appendix.

In \emph{non-collinear-safe} observables, photons arbitrarily close to an outgoing charged particle are not treated inclusively, but it is assumed that such a configuration can be experimentally resolved.
This is, for instance, the case for muons in the final state, which are observed in the muon chambers further 
out in the detector, whereas the photons are detected independently in the electromagnetic calorimeter.
Such an observable definition is sensitive to the details of the collinear $\Pf\to\Pf+\Pgg$ splitting.
The subtraction formalism can be extended to 
cover this case by exploiting the information on the dipole variable $z_{ia}$ which corresponds to the momentum fraction $z_i$ of Eq.~\eqref{eq:z_i} carried away by the emitter fermion in the collinear limit.
This is accomplished by treating the $n$-particle kinematics of the dipole phase space \lips[ia]{n} as an $(n+1)$-particle event with momenta $p_i\to z_{ia}\tilde{p}_i$ and $k\to(1-z_{ia})\tilde{p}_i$. 
For further details we refer to Ref.~\cite{Dittmaier:2008md} where this extension was introduced.
Owing to the explicit dependence of the observables on the dipole variable, the integration over $z_{ia}\equiv z$ cannot be performed analytically, but is left open for a numerical evaluation.
Employing a plus prescription, the singular endpoint part can be isolated so that the extension to the non-collinear-safe case amounts to an additional term which vanishes for inclusive observables,
\begin{align}
	\MoveEqLeft
	\int\rd\lips{n+1}\modsq{\cMsub[ia](\lips{n+1})} 
	\Theta_\cut\left(
	p_i=z_{ia}\tilde{p}_i, \,
	k=(1-z_{ia})\tilde{p}_i, \,
	\{\tilde{k}_l\} \right)
	\nonumber\\
	&= -\frac{\alpha}{2\pi}\,\sigma_iQ_i\,\sigma_aQ_a 
	\int\rd\lips[]{n} \int_0^1\rd z\,
	\biggl\{
	\Dsub{ia,\tau}(\lips[]{n}) \delta(1-z) 
	+ \left[\cDbarsub{ia,\tau}(\lips[]{n},z)\right]_+
	\biggr\}
	\nonumber\\&\quad\times
	\modsq{\cM_0(\lips[]{n};\tau\kappa_i)}
	\Theta_\cut\left(
	p_i=z\tilde{p}_i, \,
	k=(1-z)\tilde{p}_i, \,
	\{\tilde{k}_l\} \right),
\label{eq:dipole-decay:kabelschacht}
\end{align}
where the plus prescription is defined by
\begin{align}
\int_0^1\rd x\, [f(x)]_+ g(x) = \int_0^1\rd x\, f(x) [g(x)-g(1)]
\end{align}
for any smooth test function $g(x)$.
The function \cDbarsub{ia} is obtained after carrying out the analytical integration over the dipole variable $y_{ia}$ and explicitly reads
\begin{align}
	\cDbarsub{ia,+}(z) &= 
	-\left(\frac{1+z^2}{1-z}\right) \ln\left(\frac{m_i^2}{z\bar{P}_{ia}^2}[1-\eta(z)]\right)
	+(1+z)\ln(1-z) -\frac{2z}{1-z} 
	\nonumber\\*&\quad
	+(1+z)\ln\left(1+\frac{\mu_{ia}^2}{\eta(z)}\right)
	-\frac{2}{(1-z)\sigma(z)} \Biggl\{
	\ln\left(1+\frac{\eta(z)[1-z\eta(z)]}{\mu_{ia}^2(1-z)}\right)
	\nonumber\\*&\qquad
	-2\ln\left(1-\frac{2z\eta(z)}{1+\sigma(z)}\right)
	+\sigma(z)\ln\left(\frac{\mu_{ia}^2}{\eta(z)}(1-z)\right)
	\Biggr\}
	\nonumber\\*&\quad
	+\frac{2 m_a^2}{\bar{P}_{ia}^2\sigma(z)^2} \Biggr\{
	\frac{2\mu_{ia}^2(1-2z)}{\sigma(z)}
	\ln\left(\frac{2\mu_{ia}^2(1-z)+\eta(z)[1+\sigma(z)]}
	              {2\mu_{ia}^2(1-z)+\eta(z)[1-\sigma(z)]}\right)
	\nonumber\\*&\qquad	              
	-\frac{\eta(z)}{1-z[1-y_+(z)]} 
	\left( 2
	+\frac{1-\eta(z)}{\bigl(\mu_{ia}^2+\eta(z)\bigr)(1-z)} \right)
	\Biggr\} -\cDbarsub{ia,-}(z) ,
	\nonumber\\
	\cDbarsub{ia,-}(z) &=
	1-z ,
	\label{eq:dipole-decay:Dbar}
\end{align}
where $y_+(z)$ denotes the upper boundary of the $y_{ia}$ integration and is given by
\begin{align}
	y_+(z) &=
	\left[ \xi(z) +1 +\sqrt{\xi(z)[\xi(z)+2]} \right]^{-1} , &
	\xi(z) &= 
	\frac{\mu_{ia}^2}{2z(1-z)} .
\end{align}
We have further introduced the dimensionless variable
\begin{equation}
	\mu_{ia}^2 = \frac{P_{ia}^2}{\bar{P}_{ia}^2} ,
\end{equation}
and the auxiliary functions
\begin{align}
	\sigma(z) &=
	\sqrt{ 1+ 4\mu_{ia}^2 \, z(1-z) } , &
	\eta(z) &=
	\begin{cases}
	[1-y_+(z)]\, z  & \text{for } z<\frac{1}{2} , \\
	[1-y_+(z)]\, (1-z)  & \text{for } \frac{1}{2} < z  .
	\end{cases}
\end{align}


\section{The one-cutoff phase-space slicing method}
\label{sec:ocsm}

The main idea of the phase-space slicing approach is the isolation of 
resolved hard emissions from the unresolved soft and/or 
collinear regions in the real-emission phase space upon introducing a technical
cut parameter.
The hard-emission contributions are free of singularities, so that the corresponding
phase-space integration can be performed numerically using Monte Carlo methods. 
The contributions from the unresolved regions
encapsulate the IR singularities of the real corrections, 
and the integrations over the sub-spaces containing the singularities are performed analytically,
exploiting 
the universal factorization properties of amplitudes in the soft or collinear limits.
This makes the IR singularities that were implicitly contained in the 
real-emission phase space explicit by means of regulators and allows for an analytic 
cancellation of all IR singularities.
Various approaches have been proposed for the introduction of the technical cut. 
The OCS method considered in this work imposes a single cut $2p_i p_j>\Delta s$
on the invariant scalar products of the momenta $p_i,p_j$ 
of two particles that can cause a soft singularity ($p_i\to0$ or $p_j\to0$)
or a collinear singularity ($p_i p_j\to0$).
The method, thus, isolates soft and collinear singularities simultaneously and
results in a formalism that is manifestly Lorentz invariant.
The technical cut parameter $\Delta s$ should be chosen small enough to suppress
all effects of order ${\cal O}(\Delta s)$ to a numerically irrelevant level.
Despite the drawback of introducing an arbitrary resolution parameter, such as $\Delta s$,
slicing methods 
offer the possibility to widely suppress
negative weights and to avoid unbounded weights in Monte Carlo integrations,
which considerably simplifies a later extension of the Monte Carlo integrator to a
Monte Carlo event generator with unweighted events.

Although the OCS method has been extensively studied in NLO QCD calculations involving 
massless~\cite{Giele:1991vf,Giele:1993dj} or 
massive~\cite{Harris:2002md,Keller:1998tf,Cao:2008af} cases, 
there is so far no exhaustive survey of results 
covering all relevant configurations 
of massive particles in production and decay processes.
The OCS method is particularly suited to NLO QCD calculations in leading-colour
approximation where the corrections are 
decomposed into colour-ordered contributions that naturally involve only 
singularities associated with the two neighbouring hard radiators of the unresolved parton in the colour ordering.
If more than two collinear singularities appear in one contribution, as for instance in the sub-leading colour contributions,
the application of OCS becomes subtle. 
In this case, process-specific treatments are required, such as reconstructing the IR singularities from the leading-colour building blocks as described in Refs.~\cite{Giele:1991vf,Giele:1993dj}.

In the following, we describe a generic method to apply OCS 
in the presence of arbitrarily many collinear singularities in phase space.
Moreover, we are not aware of any application of 
the OCS method to QED radiation processes, 
where e.g.\ no ``colour ordering'' exists leading to amplitude contributions
with only two collinear singularities.
Furthermore, 
the results in the literature all employ dimensional regularization with either 
vanishing masses or large mass parameters; in QED applications, however, the case of small, 
but finite masses frequently occurs as well (e.g.\ for electrons or 
muons emitting photons).
It should be noted that the case of a small mass $m$ cannot be obtained upon simply 
taking the limit $m\to0$ in the OCS method, since the hierarchy
$\Delta s\ll m^2$ is used in the integration over the unresolved regions
for a radiating particle of mass $m$. 
In order to isolate the mass singularity regulated by a small mass $m$ (which defines 
a mass regularization), however, the inverted hierarchy $m^2\ll\Delta s$ is required.
In the following we describe OCS for QED radiation covering all relevant mass 
configurations (with large and small masses).
The transfer to QCD calculations is straightforward and merely 
requires the reintroduction of colour charges known from the 
massless QCD case.%
\footnote{In this paper, we consider only photon bremsstrahlung in detail,
which up to colour factors literally translates into gluon radiation 
off massless or massive quarks in QCD 
(with splittings $q\to q^*\Pg$, $q^*\to q\Pg$ and likewise for $\bar q$).
Other QCD splittings, such as
$\Pg\to\Pg^*\Pg$, $\Pg\to q^*q$, etc., either are covered by the known results
for massless quarks or do not need any special treatment for massive quarks,
because no soft or collinear singularity exists in those cases.}

\subsection{Overview of the method}
\label{sect:OCSMoverview}

The photonic real-emission corrections are divided into the finite ``hard'' contribution $I_\text{H}$
and the singular ``soft--collinear'' contribution $I_\text{SC}$ as follows,
\begin{equation}
 \int \rd\lips{n+1} \,\sum_{\lambda_\gamma}\modsq{\cM_{1}(\lips{n+1})} = I_\text{H} + I_\text{SC} .
\label{eq:OCSM}
\end{equation}
The two regions are separated with the resolution parameter $\Ds$ imposing 
conditions $s_{ f\gamma}\lessgtr\Ds$ on the Lorentz-invariant quantities $s_{ f\gamma}$ defined as
\begin{equation}
  s_{f\gamma} \equiv 2 p_f k ,
  \label{eq:s_fk}
\end{equation}
where $p_f$ and $k$ are the momenta of a charged fermion $ f$ and the photon, respectively. 
The definition~\eqref{eq:s_fk} holds irrespective whether the fermion is massless or massive.
In order to formally partition the phase space into hard and soft--collinear regions,
we introduce shorthands for the $\theta$-functions
\begin{align}
\theta_f &\equiv \theta(s_{f\gamma}-\Delta s), &
\bar\theta_f &\equiv 1-\theta_f =\theta(\Delta s-s_{f\gamma}).
\end{align}

For the \emph{hard} region we require $s_{ f\gamma}>\Ds$ for all charged fermions $f$ 
of the process, so that the corresponding contribution to the real correction reads
\begin{align}
 I_\text{H} &= \int \rd\lips{n+1}\, \sum_{\lambda_\gamma}\modsq{\cM_{1}(\lips{n+1})} \Theta , &
\qquad \Theta &= \prod_f \theta_f.
 \label{eq:OCSM_hard}
\end{align}
The phase-space integration in $I_\text{H}$ can be performed numerically, producing
a result that depends on $\Delta s$ via terms proportional to 
$\ln\Delta s$ or $\ln^2\Delta s$ for small $\Delta s$.

\begin{figure}
  \centering
  \includegraphics[width=\columnwidth]{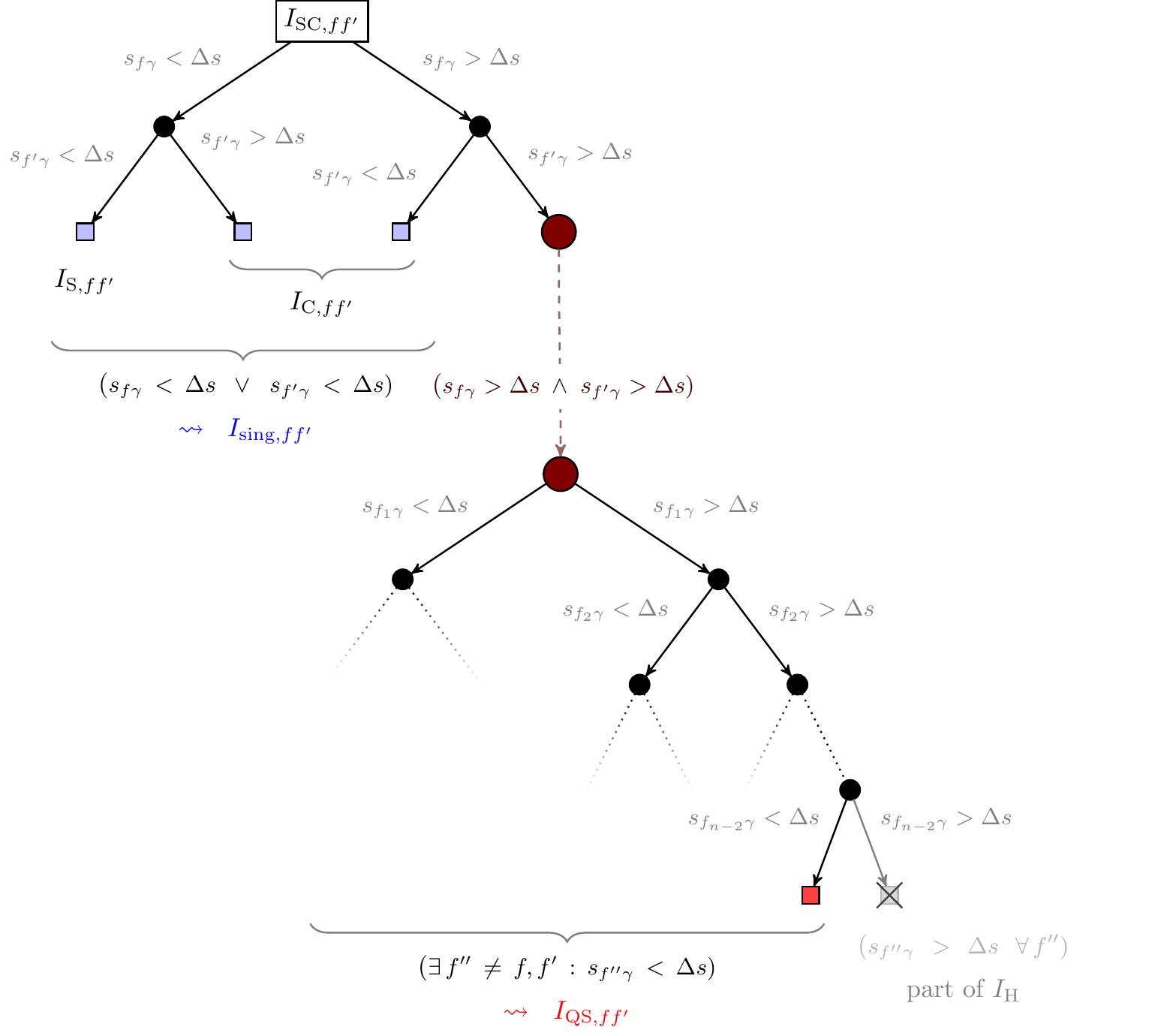}
\caption{Schematic representation of the different contributions to the soft--collinear region for a single pair of particles, $ff'$.}
\label{fig:surface_region}
\end{figure}

The integral over the \emph{soft--collinear} region is given by
\begin{equation}
  I_\text{SC} 
  = \int\rd\lips{n+1}\, \sum_{\lambda_\gamma}\modsq{\cM_{1}(\lips{n+1})} 
  \, (1-\Theta) .
  \label{eq:OCSM_SC}
\end{equation}
In this integral, at least one of the invariants $s_{f\gamma}$ is small, 
i.e.\ $s_{f\gamma}<\Delta s$.
This means that in the limit $\Delta s\to0$, non-vanishing contributions from the 
soft--collinear region to the cross section originate entirely from those terms in 
the matrix element that lead to soft or collinear singularities, because regular 
contributions tend to zero with $\Delta s\to0$. 
Thus, in the calculation of soft--collinear contributions to the cross section 
we can employ the universal factorization form of the squared amplitude, 
as for instance described in Section~\ref{sec:dipole}. In fact it is convenient to directly use
$|\M_{\sub}|^2=\sum_{f\ne f'}|\M_{\sub,ff'}|^2$
as defined in dipole subtraction in terms of emitter--spectator pairs.
We, thus, can evaluate $I_{\mathrm{SC}}$ as follows,
\begin{equation}
I_{\mathrm{SC}} = \int\rd\Phi_{n+1}\sum_{f\ne f'}|\M_{\sub,ff'}|^2(1-\Theta).
\end{equation}
For each emitter--spectator pair $ff'$ we can write $\Theta$ as
\begin{align}
\Theta &= \prod_{f''} \theta_{f''}
= \theta_f\theta_{f'}(1-\tilde\Theta_{ff'}), &
\tilde\Theta_{ff'}&=\left\{\begin{array}{ll}
1 & \mbox{if at least one $s_{f''\gamma}<\Delta s$ for $f''\ne f,f'$,} \\
0 & \mbox{otherwise}.
\end{array}\right.
\end{align}
Using $\theta_f+\bar\theta_f=1$, we can rewrite the factor
$(1-\Theta)$ to
\begin{equation}
1-\Theta = (\theta_f+\bar\theta_f)(\theta_{f'}+\bar\theta_{f'})
-\theta_f\theta_{f'}(1-\tilde\Theta_{ff'})
= \bar\theta_f\bar\theta_{f'} + \bar\theta_f\theta_{f'} + \bar\theta_{f'}\theta_f
+\theta_f\theta_{f'}\tilde\Theta_{ff'}
\end{equation}
and decompose $I_{\mathrm{SC}}$ into
soft (S), collinear (C), and quasi-soft (QS) contributions as follows,
\begin{align}
I_{\mathrm{SC}} &= \sum_{f\ne f'} I_{\mathrm{SC},ff'} =
\sum_{f\ne f'} \left(
I_{\mathrm{S},ff'} + I_{\mathrm{C},ff'} + I_{\mathrm{QS},ff'} \right),
\\
I_{\mathrm{S},ff'} &= \int\rd\Phi_{n+1}|\M_{\sub,ff'}|^2 \bar\theta_f\bar\theta_{f'},
\\
I_{\mathrm{C},ff'} &=\int\rd\Phi_{n+1}|\M_{\sub,ff'}|^2
(\bar\theta_f\theta_{f'} + \bar\theta_{f'}\theta_f),
\\
I_{\mathrm{QS},ff'} &= \int\rd\Phi_{n+1}|\M_{\sub,ff'}|^2
\theta_f\theta_{f'}\tilde\Theta_{ff'}.
\end{align}
The definition of the various sub-contributions is schematically illustrated
in Fig.~\ref{fig:surface_region}.
The integrals $I_{\mathrm{S},ff'}$ and $I_{\mathrm{C},ff'}$ are confined to the soft and
collinear regions, respectively, and involve only the technical cuts on
the invariants $s_{f\gamma}$ and $s_{f'\gamma}$ of the emitter and spectator fermions.
For this reason, the integrations in $I_{\mathrm{S},ff'}$ and $I_{\mathrm{C},ff'}$
over the singular subspaces can be performed analytically for small $\Delta s$, as outlined
below.
On the other hand, the quasi-soft integrals $I_{\mathrm{QS},ff'}$ involve cuts on
all charged particles, rendering a process-independent 
analytical treatment impossible;
nevertheless their numerical evaluation is straightforward, as outlined at the
end of this section.
Note that the quasi-soft integrals $I_{\mathrm{QS},ff'}$ do not occur
if only two sources of collinear singularities exist in the 
bremsstrahlung phase space. 
It is the occurrence of these subtle terms that 
requires process-specific and non-generic manipulations of the formalism
as described in Refs.~\cite{Giele:1991vf,Giele:1993dj} for the sub-leading colour contributions.

The analytical calculation of the singular phase-space integrals in
$I_{\mathrm{S},ff'}$ and $I_{\mathrm{C},ff'}$ starts by using the factorized
matrix element of Eq.~\eqref{eq:dipole-subtraction-general} and the
corresponding factorized $(n+1)$-particle phase space,
the construction of which is described in Ref.~\cite{Dittmaier:1999mb} in detail.
It is convenient to directly calculate the sum of the two contributions,
\begin{align}
  I_{\mathrm{sing},ff'} &= I_{\mathrm{S},ff'} + I_{\mathrm{C},ff'}
  \nonumber\\
  &= -e^2 \sigma_f Q_f\, \sigma_{f'} Q_{f'}
  \int\rd\lips[ff']{n}\otimes\int\rd[k]\,
  \gsub{ff'}(p_f,p_{f'},k)\,
  \modsq{\cM_0(\lips[ff']{n})} 
  \nonumber\\
  &\quad \times
  (\bar\theta_f\bar\theta_{f'} + \bar\theta_f\theta_{f'} + \bar\theta_{f'}\theta_f).
\label{eq:Ising}
\end{align}
In the following
we restrict ourselves to the case of unpolarized fermions with the radiator functions
\begin{equation}
  \gsub{ff'} = \gsub{ff',+} + \gsub{ff',-} .
  \label{eq:gsub_sum}
\end{equation} 
Note that the spin information of a massive decaying particle is still fully 
accessible in spite of this choice, because 
the introduction of a separate spin-flip function $\gsub{ff',-}$ was
motivated by the
massless limit for an emitter fermion, which is discontinuous.
For collinear singularities of initial-state fermions, the
multiplication symbol ``$\otimes$'' in Eq.~\eqref{eq:Ising} indicates
that a convolution over the collision energy after photon emission is involved
in the phase-space factorization. In this case, the analytical integration over
$\rd[k]$ can only be performed partially, as described in 
Sects.~\ref{subsect:finalE-initialS_vice versa} and \ref{subsect:initialE-initialS} below.
For final-state singularities, however, the squared matrix element
$\bigl\lvert{\cM_0(\lips[ff']{n})}\bigr\rvert^2$ is constant in the integration over
$\rd[k]$, so that the singular integration is entirely contained in the
universal factor
\begin{align}
  G_{ff'}(s_{ff'};\Ds)
  &= 8\pi^2 \int\rd[k] \; \gsub{ff'}(p_f,p_{f'},k) \,
(\bar\theta_f\bar\theta_{f'} + \bar\theta_f\theta_{f'} + \bar\theta_{f'}\theta_f).
\end{align}
For initial-state singularities, this factor comprises an essential ingredient as well.
Results on $G_{ff'}$ for all relevant mass configurations are provided in the
following sections.

Finally, we describe a simple way to numerically evaluate the quasi-soft
integrals $I_{\mathrm{QS},ff'}$. Recall that the integral receives only
support outside the soft--collinear region of the emitter--spectator pair $ff'$
on a phase-space volume of 
${\cal O}(\Delta s)$, because the photon
has to be collinear to at least one other charged fermion.
Non-vanishing contributions to $I_{\mathrm{QS},ff'}$ in the limit $\Delta s\to0$
can, thus, only result from an enhancement in the integrand lifting the
$\Delta s$ suppression in phase space. Power-counting in the photon momentum
in the integrand reveals that this can only happen if the photon momentum 
becomes soft, i.e.\ the photon momentum has to be of order ${\cal O}(\Delta s)$
in order to lie outside the soft--collinear region of $ff'$.
The integral $I_{\mathrm{QS},ff'}$ is therefore 
effectively a surface integral
of the soft--collinear region of $ff'$ and can be evaluated upon taking into
account only those parts in the integrand that produce soft singularities, i.e.\
the ``eikonal terms'' in $\gsub{ff'}(p_f,p_{f'},k)$ are sufficient,
\begin{equation}
\gsubeik{ff'}(p_f,p_{f'},k) =
\frac{1}{p_f k} \left[ \frac{2p_f p_{f'}}{p_f k+p_{f'}k}-\frac{m_f^2}{p_f k} \right].
\end{equation}
After symmetrizing with respect to the interchange $f\leftrightarrow f'$,
we can identify the usual eikonal factor
\begin{align}
\geik{ff'}(p_f,p_{f'},k) &=  \frac{1}{2}
\left[\gsubeik{ff'}(p_f,p_{f'},k)+\gsubeik{f'f}(p_{f'},p_f,k) \right]
\nonumber\\
&=
\frac{p_f p_{f'}}{(p_f k)(p_{f'} k)}  
-\frac{m_f^2}{2(p_f k)^2} 
-\frac{m_{f'}^2}{2(p_{f'} k)^2}.
\end{align}
Since only the symmetric combination of $\gsub{ff'}(p_f,p_{f'},k)$ is relevant
in the overall quasi-soft contribution (summed over all emitter--spectator
combinations), $I_{\mathrm{QS},ff'}$ can be evaluated via
\begin{equation}
I_{\mathrm{QS},ff'} = 
  -e^2 \,\sigma_f Q_f\, \sigma_{f'} Q_{f'}
\int\rd\Phi_{n+1}\,
  \geik{ff'}(p_f,p_{f'},k)\,
  \modsq{\cM_0(\lips[ff']{n})} \,
\theta_f\theta_{f'}\tilde\Theta_{ff'}.
\label{eq:IQSff}
\end{equation}
Note that this form is valid for any combination of incoming and outgoing
emitters and spectators.
In the form given in Eq.~\eqref{eq:IQSff}, each term $I_{\mathrm{QS},ff'}$
involves the phase-space embedding $\Phi_{n+1}\to \lips[ff']{n}$ 
that is specific to each emitter--spectator pair. 
Since, however, the integral receives contributions only from soft photon
momenta $k$, we can simplify the evaluation even further.
For the relevant contribution we can factorize the $(n+1)$-particle
phase space into the corresponding $n$-particle phase space for all
particles but the photon and the 1-particle photon phase space,
leading to the alternative form
\begin{equation}
I_{\mathrm{QS},ff'} = 
  -e^2 \sigma_f Q_f\, \sigma_{f'} Q_{f'}
\int\rd\Phi_{n}\,
  \modsq{\cM_0(\Phi_{n})} \,
\int\frac{\rd^3{\bf k}}{(2\pi)^32k^0} \,
  \geik{ff'}(p_f,p_{f'},k)\,
\theta_f\theta_{f'}\tilde\Theta_{ff'},
\end{equation}
which does not make use of any special phase-space mapping for
$ff'$ pairs and is, thus, much simpler to evaluate than Eq.~\eqref{eq:IQSff}.
In this approximation, some care has to be taken with respect to the upper bound 
of the photon energy $k^0$, which has to be large enough to cover the full
relevant soft region, but still small enough to avoid artefacts from 
terms of the order ${\cal O}(\Delta s \ln k^0_\max)$.

In the following section we carry out the integrations over the 
soft--collinear regions in $I_{\mathrm{sing},ff'}$ for the various
emitter--spectator pairs $ff'$ in the initial/final states.

\subsection{Integration over the singular regions}
\label{sect:OCSM_IR}

\subsubsection{Final-state emitter and final-state spectator}
\label{subsect:final-final}

%
\begin{figure}
  \centering
  \includegraphics[scale=1.25]{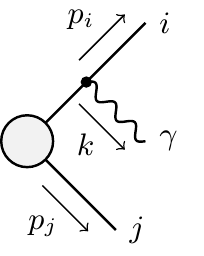}
\caption{Generic diagram with a final-state emitter $i$ and a final-state spectator $j$.}
\label{fig:final-final}
\end{figure}
Following the notation and conventions of Refs.~\cite{Dittmaier:1999mb,Dittmaier:2008md},
in the case of a final-state emitter $i$
and a final-state spectator $j$ 
(illustrated in Fig.~\ref{fig:final-final})
the full integral over the soft--collinear regions is given by
\begin{equation}
 G_{ij}(P_{ij}^2,\Delta s)=\frac{\PijB^4}{2\sqrt{\lambda_{ij}}}\int_{y_1}^{y_2}\rd y_{ij}\,
(1-y_{ij})\int_{z_1(y_{ij})}^{z_2(y_{ij})}\rd z_{ij}\,
\gsub{ij}(p_i,p_j,k)\,
(\bar\theta_i\bar\theta_j + \bar\theta_i\theta_j + \theta_i\bar\theta_j),
  \label{eq:Gij_ff_gen}
\end{equation}
where
\begin{align}
 P_{ij}&=p_i+p_j+k, & 
 \PijB^2&=P_{ij}^2-m_i^2-m_j^2-m_{\gamma}^2, & 
\lambda_{ij}&=\lambda(P_{ij}^2,m_i^2,m_j^2),
\end{align}
with $\lambda(a,b,c)$ defined in Eq.~(\ref{eq:kallen}).
The variables
\begin{align}
y_{ij} &= \frac{p_i k}{p_i p_j + p_i k + p_j k}, &
z_{ij} &= \frac{p_i p_j}{p_i p_j + p_j k}
\end{align}
parametrize the photon phase space. Their boundaries are given by
\begin{align}
 y_1 &= \frac{2m_im_\gamma}{\PijB^2},\quad 
y_2=1-\frac{2m_j\left(\sqrt{P_{ij}^2}-m_j\right)}{\PijB^2}, \notag \\
z_{1,2}(y)&=\frac{(1-y)(2m_i^2+\PijB^2 y)\mp\sqrt{y^2-y_1^2}\sqrt{\lambda_{ij}}R_{ij}(y)}{2(1-y)(m_i^2+m_\gamma^2+\PijB^2 y)},
\label{eq:yz_ff}
\end{align}
where
\begin{equation}
 R_{ij}(y)=\,\frac{\sqrt{(2m_j^2+\PijB^2-\PijB^2 y)^2-4P_{ij}^2m_j^2}}{\sqrt{\lambda_{ij}}}.
\end{equation}
Note that the infinitesimal photon mass $m_\gamma$ acts as regulator for the
soft singularity appearing at $y_{ij}\to0$.
The integrand $\gsub{ij}(p_i,p_j,k)$ is given by
\begin{equation}
\gsub{ij}(p_i,p_j,k) =
\frac{1}{(p_i k)R_{ij}(y_{ij})} \biggl[
\frac{2}{1-z_{ij}(1-y_{ij})}-1-z_{ij}
-\frac{m_i^2}{p_i k} \biggr]
\end{equation}
and the soft and collinear regions are characterized as follows,
\begin{subequations}
\label{eqs:finfin}
\begin{align}
  \label{eq:finfin_softcond}
  &\text{soft region:} & 
  k & \to 0 &
  \begin{cases}
  s_{i\gamma}<\Ds &\Rightarrow\; y_{ij}<\Delta y , \\
  s_{j\gamma}<\Ds &\Rightarrow\; z_{ij}>1-\Delta z , 
  \end{cases}
  \\[.5em]
  \label{eq:finfin_coll1cond}
  &\text{$i$-collinear region:} &
  k&\nrightarrow0, \quad k \parallel p_i &
  \begin{cases}
  s_{i\gamma}<\Ds &\Rightarrow\; y_{ij}<\Delta y , \\
  s_{j\gamma}>\Ds &\Rightarrow\; z_{ij}<1-\Delta z , 
  \end{cases}
  \\[.5em]
  \label{eq:finfin_coll2cond}
  &\text{$j$-collinear region:} &
  k&\nrightarrow0, \quad k \parallel p_j &
  \begin{cases}
  s_{i\gamma}>\Ds &\Rightarrow\; y_{ij}>\Delta y , \\
  s_{j\gamma}<\Ds &\Rightarrow\; z_{ij}>1-\Delta z , 
  \end{cases}
\end{align}
\end{subequations}
where
\begin{align}
 \Delta y&=\frac{\Ds}{\PijB^2}, & \Delta z&=\frac{\Delta y}{1-y_{ij}}.
\end{align}
Depending on the masses of the final-state particles (massive or light emitter/spectator), 
we obtain different results for $G_{ij}(P_{ij}^2,\Delta s)$, which are listed below.

The full integral over the soft--collinear regions, as defined in Eq.~\eqref{eq:Gij_ff_gen},
is sufficient to cover all kinematic configurations in which the photon is
recombined with collinear fermions in the event selection, i.e.\
in the case of collinear-safe observables.
If the photon and the emitter $i$ become collinear, this photon recombination 
ensures that all 
$i$-collinear configurations integrated over in $G_{ij}(P_{ij}^2,\Delta s)$
are treated inclusively, i.e.\ either the full contribution represented by
$G_{ij}(P_{ij}^2,\Delta s)$ contributes or it is completely excluded by cuts.
This even holds true for contributions in $G_{ij}(P_{ij}^2,\Delta s)$ where
the photon is collinear to the spectator $j$, since the photon has to be soft in this case,
i.e.\ $s_{j\gamma}={\cal O}(\Delta s)$.
In non-collinear-safe observables,
however, the photon can be separated from collinear fermions,
which is for instance the case for muons detected in the muon chambers.
In this case
the differential information on the energy flow inside the $i$-collinear cone
should be kept in the phase-space integral \eqref{eq:Gij_ff_gen}.
This energy flow in the soft--collinear region is controlled by the variable $z_{ij}$,
where it is effectively given by $z_i=p_i^0/(p_i^0+k^0)$.
In this situation, the integration over $z_{ij}$ in Eq.~\eqref{eq:Gij_ff_gen}
should be included in the numerical phase-space integration, so that instead
of the full integral $G_{ij}(P_{ij}^2,\Delta s)$ the following function of 
$z=z_{ij}$ appears 
\begin{equation}
\bar{\cG}_{ij}(P_{ij}^2,\Delta s,z)
=\frac{\PijB^4}{2\sqrt{\lambda_{ij}}}\int_{y_1(z)}^{y_2(z)}\rd y_{ij}\,
(1-y_{ij})\,
\gsub{ij}(p_i,p_j,k)\,
(\bar\theta_i\bar\theta_j + \bar\theta_i\theta_j + \bar\theta_j\theta_i),
\label{eq:cGij}
\end{equation}
where the functions $y_{1,2}(z)$ result from inverting the parametrization
of the integral boundary defined in Eq.~\eqref{eq:yz_ff}.
The evaluation of $\bar{\cG}_{ij}(P_{ij}^2,\Delta s,z)$
essentially follows the analogous treatment described for dipole
subtraction in the case of non-collinear observables in Ref.~\cite{Dittmaier:2008md}.
The procedure is greatly simplified upon introducing a plus distribution for
the $z$-integration,
\begin{equation}
\bar{\cG}_{ij}(P_{ij}^2,\Delta s,z) =
\left[\bar{\cG}_{ij}(P_{ij}^2,\Delta s,z)\right]_+
+ \delta(1-z) G_{ij}(P_{ij}^2,\Delta s),
\end{equation}
which isolates the soft singularity in the endpoint function $G_{ij}(P_{ij}^2,\Delta s)$,
so that $\bigl[\bar{\cG}_{ij}(P_{ij}^2,\Delta s,z)\bigr]_+$ can be
evaluated for $m_\gamma=0$.
The explicit evaluation of $\bar{\cG}_{ij}(P_{ij}^2,\Delta s,z)$ and
of the full integral $G_{ij}(P_{ij}^2,\Delta s)$ requires a discrimination
of the different mass patterns in the emitter--spectator pairs.

\paragraph{Massive emitter and massive spectator}

For massive emitter and spectator fermions, we have the following hierarchy of
parameters,
\begin{equation}
m_\gamma^2 \ll \Delta s \ll m_i^2,m_j^2,P_{ij}^2.
\label{eq:massiveijhierarchy}
\end{equation}
The full integral $G_{ij}(P_{ij}^2,\Delta s)$ is tedious, but straightforward.
The result involves a discrimination between the variables $m_i^2,m_j^2,P_{ij}^2$
and is given by
\begin{align}
G_{ij}(P_{ij}^2,\Delta s) &=
\frac{\PijB^2}{\sqrt{\lambda_{ij}}}\left[
\ln a_1\ln\left(\frac{m_\gamma^2\lambda_{ij}}{\Delta s^2 P_{ij}^2}\right)
+2\Li_{2}(a_1)-\frac{\pi^2}{3}+\frac{1}{2}\ln^2 a_1 \right]
-2\ln\left(\frac{\Delta s}{m_\gamma m_i}\right)+2
\nonumber\\
&\quad
+ f_{ij,+}(P_{ij}^2) \;\mp\; \theta(m_i^2+m_j^2-\PijB^2)\,f_{ij,\pm}(P_{ij}^2)
\qquad \mbox{for} \quad 
m_i \lessgtr m_j,
\end{align}
with the auxiliary functions
\begin{align}
f_{ij,\pm}(P_{ij}^2) = \pm\frac{\PijB^2}{\sqrt{\lambda_{ij}}}\Biggl[
& 2\ln\left(\frac{\PijB^2\pm\sqrt{\lambda_{ij}}+2m_j^2}{\PijB^2\pm\sqrt{\lambda_{ij}}}\right)
\ln\left(\frac{2m_j^2}{\PijB^2\pm\sqrt{\lambda_{ij}}}\right)
+2\Li_{2}\left(-\frac{2m_j^2}{\PijB^2\pm\sqrt{\lambda_{ij}}}\right)
\nonumber\\
&{} +\frac{\pi^2}{6}
-\frac{\PijB^2\mp\sqrt{\lambda_{ij}}}{\PijB^2}
\ln\left(\frac{2m_j^2}{\PijB^2\pm\sqrt{\lambda_{ij}}}\right)
+1\mp\frac{\sqrt{\lambda_{ij}}}{\PijB^2}
-\frac{2m_i^2}{\PijB^2}
\Biggr]
\end{align}
and the variable
\begin{equation}
a_1 = \frac{\bar P_{ij}^2+2m_i^2-\sqrt{\lambda_{ij}}}
{\bar P_{ij}^2+2m_i^2+\sqrt{\lambda_{ij}}}.
\end{equation}

To calculate the function $\bar{\cG}_{ij}(P_{ij}^2,\Delta s,z)$ for non-collinear-safe
observables, we have to invert the functions $z_{1,2}(y)$ of Eq.~\eqref{eq:yz_ff}
using the hierarchy \eqref{eq:massiveijhierarchy}. 
Making use of the fact that only $y_{ij}$ values of 
${\cal O}(\Delta y)$ with
$\Delta y\ll1$ are relevant,
the functions $y_{1,2}(z)$ are given by
\begin{align}
y_1(z) &= \frac{2m_i^2}{\PijB^2-\sqrt{\lambda_{ij}}}\,(1-z), &
y_2(z) &= \frac{2m_i^2}{\PijB^2+\sqrt{\lambda_{ij}}}\,(1-z).
\end{align}
This shows that for $m_i,m_j$ obeying Eq.~\eqref{eq:massiveijhierarchy}
all $z$ values in the soft--collinear region are of the order
$1-{\cal O}(\Delta y)$. The integral $G_{ij}(P_{ij}^2,\Delta s)$, thus,
only receives contributions from a tiny neighbourhood of the point
$(y_{ij},z_{ij})=(0,1)$, i.e.\ it does not
involve an integration over the momentum flow in the collinear regions.
In this case non-collinear-safe observable can be calculated using
$G_{ij}(P_{ij}^2,\Delta s)$ without the need to introduce
$\bar{\cG}_{ij}(P_{ij}^2,\Delta s,z)$.

\paragraph{Massive emitter $i$ and light spectator $j$}

Due to the small mass of the spectator, the hierarchy between the parameters is 
\begin{equation}
 m_\gamma^2\ll \Ds \ll m_i^2,P_{ij}^2, \qquad m_j=0.
\end{equation}
The spectator mass can be set to zero, since no collinear singularity exists
for the spectator.
In fact the integral $G_{ij}(P_{ij}^2,\Delta s)$ can be obtained upon
taking the limit $m_j\to0$ from the result of the previous section,
\begin{align}
G_{ij}(P_{ij}^2,\Delta s) &=
{\cal L}(P_{ij}^2,m_i^2)
-2\ln\biggl(\frac{\Ds}{\PijB^2}\biggr)\ln\left(\frac{m_i^2}{P_{ij}^2}\right)
-2\ln\biggl(\frac{\Ds}{\PijB^2}\biggr)
-\frac{3}{2}\ln\left(\frac{m_i^2}{P_{ij}^2}\right)
\nonumber\\
&\quad
+ \theta(\PijB^2-m_i^2) \left[ 
2\Li_{2}\left(\frac{m_i^2}{P_{ij}^2}\right)-\frac{\pi^2}{6}
+2-\frac{2m_i^2}{\PijB^2}
+\ln^2\left(\frac{m_i^2}{P_{ij}^2}\right)
-2\ln\left(\frac{\PijB^2}{m_i^2}\right)
\right] ,
\end{align}
with the function ${\cal L}$ from Eq.~\eqref{eq:L}.
Similarly to the case of non-vanishing $m_j$, 
$G_{ij}(P_{ij}^2,\Delta s)$ does not comprise an integral over the
energy flow in the collinear regions, so that the quantity 
$\bar{\cG}_{ij}(P_{ij}^2,\Delta s,z)$ is not needed either.

\paragraph{Light emitter $i$ and massive spectator $j$}

For a light emitter with mass $m_i$ we have to consider the hierarchy
\begin{equation}
m_\gamma^2\ll m_i^2\ll \Ds \ll m_j^2,P_{ij}^2. 
\end{equation}
Since the two limits $m_i^2\ll \Ds$ and $\Ds\ll m_i^2$ do not commute,
the integral $G_{ij}(P_{ij}^2,\Delta s)$ does not result from the simple
limit $m_i\to0$ applied to the previous cases.
An explicit integration delivers the result
\begin{align}
  G_{ij}(P_{ij}^2,\Ds) &=
{\cal L}\left(\frac{\PijB^4}{P_{ij}^2},m_i^2\right)
-\ln^2\left(\frac{\Ds P_{ij}^2}{\PijB^4}\right) 
-\frac{3}{2}\ln\left(\frac{\Ds P_{ij}^2}{\PijB^4}\right) 
-\frac{2\pi^2}{3}
+\frac{3}{2}
\nonumber \\
&\quad +\theta(\PijB^2-m_j^2)
\left[\frac{\pi^2}{6}+2\Li_{2}\left(-\frac{m_j^2}{\PijB^2}\right) 
+2\ln\left(\frac{m_j^2}{\PijB^2}\right)\ln\left(\frac{P_{ij}^2}{\PijB^2}\right)\right].
\label{eq:Gij_fin}
\end{align}
It contains a term $\ln^2\Ds$ that results from a combination of the
soft and the $i$-collinear singularities, which are both isolated by the 
cut parameter $\Ds$ in the integration over the hard photon emission which
is based on matrix elements with $m_i=0$.

In order to calculate $\bar{\cG}_{ij}(P_{ij}^2,\Delta s,z)$ for non-collinear-safe
observables, we first derive the functions $y_{1,2}(z)$
for the integration boundary of the $y_{ij}$-integration in Eq.~\eqref{eq:cGij}.
The relevant range in $y_{ij}$ is characterized by
${\cal O}(m_i^2/\PijB^2)\lesssim y_{ij} \lesssim {\cal O}(\Delta y)$, leading to
the region defined by
\begin{align}
y_1(z) &= \frac{m_i^2}{\PijB^2}\,\frac{1-z}{z}, &
y_2(z) &= \Delta y.
\end{align}
The function $\bar{\cG}_{ij}(P_{ij}^2,\Delta s,z)$ for
the non-collinear-safe contribution is easily calculated to
\begin{equation}
 \bar{\cG}_{ij}(P_{ij}^2,\Delta s,z)
=\frac{1+z^2}{1-z}\ln\left(\frac{\Ds z}{m_i^2 (1-z)}\right)-\frac{2z}{1-z}.
\label{eq:cGijexplicit}
\end{equation}

\paragraph{Light emitter $i$ and light spectator $j$}

In this case $G_{ij}(P_{ij}^2,\Ds)$ 
can be calculated upon
taking the massless limit $m_j \to 0$ of the previous case. The final result is
\begin{align}
G_{ij}(P_{ij}^2,\Ds) &= 
{\cal L}(P_{ij}^2,m_i^2)
-\ln^2\left(\frac{\Ds}{P_{ij}^2}\right)
-\frac{3}{2}\ln\left(\frac{\Ds}{P_{ij}^2}\right)
-\frac{\pi^2}{2}
+\frac{3}{2},
\label{eq:Gij}
\end{align}
with $\bar{\cG}_{ij}(P_{ij}^2,\Delta s,z)$ as given in Eq.~\eqref{eq:cGijexplicit}.

\subsubsection{Final-state emitter and initial-state spectator, and vice versa}
\label{subsect:finalE-initialS_vice versa}

\begin{figure}
\begin{minipage}{.45\columnwidth}
  \centering
  \includegraphics[scale=1.25]{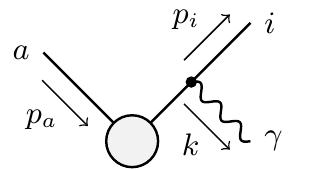}
\end{minipage}
\hfill
\begin{minipage}{.45\columnwidth}
  \centering
  \includegraphics[scale=1.25]{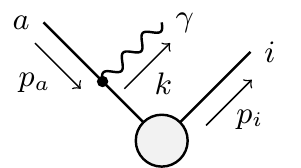}
\end{minipage}
\caption{Generic diagrams with a final-state emitter $i$ and an initial-state spectator $a$~(left), and with an initial-state emitter $a$ and a final-state spectator $i$~(right).}
\label{fig:finE-initS}
\end{figure}

In the cases of a final-state emitter $i$ and an initial-state spectator $a$, 
and of an initial-state emitter $a$ and a final-state
spectator $i$ (depicted on the left- and on the right-hand sides of 
Fig.~\ref{fig:finE-initS}, respectively), again the variables and 
abbreviations of Ref.~\cite{Dittmaier:1999mb} are used.
The integrals over the full soft--collinear regions are defined as
\begin{align}
 G_{ia}(P_{ia}^2,\Delta s) &=\,
\int_{x_0}^{x_1}\rd x_{ia}\,\frac{\bar P_{ia}^4}{2\sqrt{\lambda_{ia}}R_{ia}(x_{ia})}
\int_{z_1(x_{ia})}^{z_2(x_{ia})}\rd z_{ia}\, 
 \gsub{ia}(p_i,p_a,k)\,
(\bar\theta_i\bar\theta_a + \bar\theta_i\theta_a + \theta_i\bar\theta_a),
  \label{eq:Gia}
\\
 G_{ai}(P_{ia}^2,\Delta s) &=\,
\int_{x_0}^{x_1}\rd x_{ia}\,\frac{\bar P_{ia}^4}{2\sqrt{\lambda_{ia}}R_{ia}(x_{ia})}
\int_{z_1(x_{ia})}^{z_2(x_{ia})}\rd z_{ia}\, 
 \gsub{ai}(p_a,p_i,k)\,
(\bar\theta_i\bar\theta_a + \bar\theta_i\theta_a + \theta_i\bar\theta_a),
  \label{eq:Gai}
\end{align}
where the variables
\begin{align}
x_{ia} &= \frac{p_a p_i + p_a k - p_i k}{p_a p_i + p_a k}, &
z_{ia} &= \frac{p_a p_i}{p_a p_i + p_a k}.
\label{eq:xiazia}
\end{align}
are confined to the boundary determined by the variables
\begin{align}
z_{1,2}(x) &=
\frac{\bar P_{ia}^2[\bar P_{ia}^2-x(\bar P_{ia}^2+2m_i^2)] \mp
\sqrt{\bar P_{ia}^4(1-x)^2-4m_i^2 m_\gamma^2 x^2}\,
\sqrt{\lambda_{ia}}\, R_{ia}(x) }
{2\bar P_{ia}^2[\bar P_{ia}^2-x(P_{ia}^2-m_a^2)]}, \qquad 
\nonumber\\
x_1 &= \displaystyle{1+\frac{2m_im_\gamma}{\bar P_{ia}^2}},
\end{align}
with $x_0$ being set by the least energy to trigger the process without photon emission.%
\footnote{For $P_{ia}^2>0$ and $0 < \sqrt{P_{ia}^2} < m_a-m_i$, the lower limit
has to obey $x_0 > \hat x = \frac{-\bar P_{ia}^2}{2m_a\left(m_a-\sqrt{P_{ia}^2}\right)}$,
see Ref.~\cite{Dittmaier:1999mb}.}
Here we use 
\begin{align}
P_{ia}&=p_i+k-p_a, &
\bar P_{ia}^2 &= P_{ia}^2-m_a^2-m_i^2-m_\gamma^2, &
\lambda_{ia}&=\lambda(P_{ia}^2,m_i^2,m_a^2)
\end{align}
with $\lambda(a,b,c)$ defined in Eq.~(\ref{eq:kallen}) and
\begin{align}
 R_{ia}(x)=&\,\frac{\sqrt{(\bar P_{ia}^2+2m_a^2x)^2-4P_{ia}^2m_a^2x^2}}{\sqrt{\lambda_{ia}}}.  
\end{align}
Note that we always have $\bar P_{ia}^2<0$, but $P_{ia}^2$ is not necessarily negative for
massive particles $i$ and $a$.
The soft and collinear regions are characterized as follows,
\begin{subequations}
\label{eqs:finEinitS}
\begin{align}
  \label{eq:finEinitS_softcond}
  &\text{soft region:} & 
  k & \to 0 &
  \begin{cases}
  s_{i\gamma}<\Ds &\Rightarrow\; x_{ia} > (1+\Delta x)^{-1} , \\
  s_{a\gamma}<\Ds &\Rightarrow\; z_{ia} > 1-\Delta z ,
  \end{cases}
  \\[.5em]
  \label{eq:finEinitS_coll1cond}
  &\text{$i$-collinear region:} &
  k&\nrightarrow0, \quad k \parallel p_i &
  \begin{cases}
  s_{i\gamma}<\Ds &\Rightarrow\; x_{ia} > (1+\Delta x)^{-1} , \\
  s_{a\gamma}>\Ds &\Rightarrow\; z_{ia} < 1-\Delta z ,
  \end{cases}
  \\[.5em]
  \label{eq:finEinitS_coll2cond}
  &\text{$a$-collinear region:} &
  k&\nrightarrow0, \quad k \parallel p_a &
  \begin{cases}
  s_{i\gamma}>\Ds &\Rightarrow\; x_{ia} < (1+\Delta x)^{-1} , \\
  s_{a\gamma}<\Ds &\Rightarrow\; z_{ia} > 1-\Delta z ,
  \end{cases}
\end{align}
\end{subequations}
where
\begin{align}
\Delta x &= \frac{\Ds}{-\bar P_{ia}^2}, &  
\Delta z &= x_{ia}\Delta x.
\end{align}
By construction, the limits $x_{0,1}$ respect the condition $x_0<1/(1+\Delta x)<x_1.$
Results for the integrals $G_{ia}(P_{ia}^2,\Delta s)$ and $G_{ai}(P_{ia}^2,\Delta s)$ 
for the various mass configurations of emitter--spectator pairs are listed in the following sections.

Similar to the previous case of both emitter and spectator in the final state,
the full integrals \eqref{eq:Gia} and \eqref{eq:Gai} are not sufficient to
deal with non-collinear-safe observables if the emitter particle $f=i$ or $f=a$ is light,
where $m_f^2\ll \Delta s$.
To keep track of the momentum flow in the collinear region of the emitter,
the $z_{ia}$ or the $x_{ia}$ integration has to be done numerically in the
respective cases. To this end, we define
\begin{align}
\bar\cG_{ia}(P_{ia}^2,\Delta s,z) &=\,
\int_{x_1(z)}^{x_2(z)}\rd x_{ia}\,\frac{\bar P_{ia}^4}{2\sqrt{\lambda_{ia}}R_{ia}(x_{ia})}\,
 \gsub{ia}(p_i,p_a,k)\,
(\bar\theta_i\bar\theta_a + \bar\theta_i\theta_a + \theta_i\bar\theta_a),
  \label{eq:cGia}
\\
\cG_{ai}(P_{ia}^2,\Delta s,x) &=\,
\frac{\bar P_{ia}^4}{2\sqrt{\lambda_{ia}}R_{ia}(x)}
\int_{z_1(x)}^{z_2(x)}\rd z_{ia}\, 
 \gsub{ai}(p_a,p_i,k)\,
(\bar\theta_i\bar\theta_a + \bar\theta_i\theta_a + \theta_i\bar\theta_a),
  \label{eq:cGai}
\end{align}
using $x=x_{ia}$ and $z=z_{ia}$ as shorthands.
In the case of a final-state emitter, $zp_i$ is the fermion momentum after collinear
photon emission from the final state;
in the case of an initial-state emitter, $xp_a$ is the fermion momentum after collinear
photon emission from the initial state, i.e.\ the momentum $x p_a$ enters the hard
non-radiative process.
The functions $x_{1,2}(z)$ appearing in Eq.~\eqref{eq:cGia} 
result from interchanging the order in the integrations
over $x_{ia}$ and $z_{ia}$ and are derived below for the relevant cases.
The soft singularities in the integrals again are extracted from the numerical integration upon
introducing plus distributions in the integrals,
\begin{align}
\bar{\cG}_{ia}(P_{ia}^2,\Delta s,z) &=
\left[\bar{\cG}_{ia}(P_{ia}^2,\Delta s,z)\right]_+
+ \delta(1-z) G_{ia}(P_{ia}^2,\Delta s),
\\
{\cG}_{ai}(P_{ia}^2,\Delta s,x) &=
\left[{\cG}_{ai}(P_{ia}^2,\Delta s,x)\right]_+
+ \delta(1-x) G_{ai}(P_{ia}^2,\Delta s).
\end{align}

\paragraph{Massive final-state emitter $i$ and massive initial-state spectator $a$, and vice versa}

The case of massive initial-state particles is mostly interesting for decay processes,
since the masses of 
colliding particles are typically much smaller than
the scattering energies in high-energy collisions.
If both final-state emitter $i$ and initial-state spectator $a$ (or vice versa)
are massive, we have to respect the following hierarchy of parameters,
\begin{equation}
m_\gamma^2 \ll \Delta s \ll m_i^2,m_a^2,|P_{ia}^2|,|\bar P_{ia}^2|.
\label{eq:massiveiahierarchy}
\end{equation}
No collinear singularities are encountered, and the effective integration
region in $G_{ia,ai}(P_{ia}^2,\Delta s)$ is a neighbourhood of the point
$(x_{ia},z_{ia})=(1,1)$, so that the functions $\bar{\cG}_{ia}(P_{ia}^2,\Delta s,z)$
and ${\cG}_{ai}(P_{ia}^2,\Delta s,x)$ are not required.
The full integrals read
\begin{align}
G_{ia}(P_{ia}^2,\Delta s) &=
\frac{\bar P_{ia}^2}{\sqrt{\lambda_{ia}}}\left\{
\ln(b_1)\ln\left(\frac{\Ds^2 (m_a^2+m_i^2-\bar P^2_{ia})}{m_\gamma^2 \lambda_{ia}}\right)
-2\Li_{2}(b_1)
+\frac{\pi^2}{3}-\frac{1}{2}\ln^2(b_1)
\right\}
\nonumber\\
&\quad
+2\ln\left(\frac{m_\gamma m_i}{\Ds}\right)+2 
+ f_{ia,+}(P_{ia}^2) \;\mp\; \theta(P_{ia}^2)\,f_{ia,\pm}(P_{ia}^2)
\qquad \mbox{for} \quad m_i \lessgtr m_a,
  \label{eq:Gia2}
\\
G_{ai}(P_{ia}^2,\Delta s) &=   
\frac{\bar P_{ia}^2}{\sqrt{\lambda_{ia}}} \Biggl\{
\ln\left(\frac{b_1}{c_0}\right)\ln\left(\frac{m_\gamma^2 \lambda_{ia}}{m_a^2\Ds^2}\right)
+\frac{1}{2}\ln({c_0b_1})\ln\left(\frac{b_1}{c_0}\right)
\nonumber \\
& \hspace*{4em} {}
-\ln\left(\frac{m_a^2+m_i^2-\bar P_{ia}^2}{m_a^2}\right)\ln(b_1)
-2\Li_{2}(c_0)+2\Li_{2}(b_1)+\ln(c_0)
\Biggr\}
\nonumber\\
&\quad
+2\ln\left(\frac{m_\gamma m_i}{\Ds}\right)
+ f_{ai,+}(P_{ia}^2) \;\mp\; \theta(P_{ia}^2)\,f_{ai,\pm}(P_{ia}^2)
\qquad \mbox{for} \quad m_i \lessgtr m_a,
  \label{eq:Gai2}
\end{align}
with the auxiliary functions
\begin{align}
f_{ia,\pm}(P_{ia}^2)  &=
\pm\frac{\bar P_{ia}^2}{\sqrt{\lambda_{ia}}}\Biggl\{
-\ln^2\left(\frac{2m_i^2}{2m_i^2-\bar P_{ia}^2\mp\sqrt{\lambda_{ia}}}\right)
-2\Li_{2}\left(\frac{2m_i^2}{2m_i^2-\bar P_{ia}^2\mp\sqrt{\lambda_{ia}}}\right)
+\frac{\pi^2}{6}
\notag \\
& \hspace*{5em} {}
-\frac{\bar P_{ia}^2\pm\sqrt{\lambda_{ia}}}{\bar P_{ia}^2}
\ln\left(\frac{2m_i^2}{-\bar P_{ia}^2\mp\sqrt{\lambda_{ia}}}\right)
-1-\frac{2m_i^2}{\bar P_{ia}^2} \mp \frac{\sqrt{\lambda_{ia}}}{\bar P_{ia}^2} \Biggr\},
\\
f_{ai,\pm}(P_{ia}^2)  &=
\pm\frac{\bar P_{ia}^2}{\sqrt{\lambda_{ia}}}\Biggl\{
2\ln\left(\frac{-\bar P_{ia}^2\mp\sqrt{\lambda_{ia}}}{2m_i^2-\bar P_{ia}^2\mp\sqrt{\lambda_{ia}}}\right)
\ln\left(\frac{2m_i^2}{-\bar P_{ia}^2\mp\sqrt{\lambda_{ia}}}\right)
-2\Li_{2}\left(\frac{2m_i^2}{\bar P_{ia}^2\pm\sqrt{\lambda_{ia}}}\right)
\notag \\
& \hspace*{5em} {}
-\frac{\pi^2}{6}
+\frac{\bar P_{ia}^2\mp\sqrt{\lambda_{ia}}}{\bar P_{ia}^2}
\ln\left(\frac{2m_i^2}{-\bar P_{ia}^2\mp\sqrt{\lambda_{ia}}}\right)
-1-\frac{2m_a^2}{\bar P_{ia}^2} \pm \frac{\sqrt{\lambda_{ia}}}{\bar P_{ia}^2} \Biggr\}.
\end{align}
Here we 
partially used the results of Ref.~\cite{Dittmaier:1999mb}, 
where we have identified $(1-x_0)$ with $\Delta x$
in the limit $\Delta x\rightarrow0$, and 
\begin{align}
 b_1&=\frac{2m_i^2-\bar P_{ia}^2-\sqrt{\lambda_{ia}}}{2m_i^2-\bar P_{ia}^2+\sqrt{\lambda_{ia}}}, & 
 c_0&=\frac{\bar P_{ia}^2+\sqrt{\lambda_{ia}}}{\bar P_{ia}^2 -\sqrt{\lambda_{ia}}}.
\label{eq:b_1c_0}
\end{align}

\paragraph{Light particle $i$ and massive particle $a$}

If particle $i$ is light and particle $a$ is heavy, 
we have to consider the hierarchy 
\begin{equation}
m_\gamma^2\ll m_i^2 \ll \Ds \ll m_a^2,|P_{ia}^2|,|\bar P_{ia}^2|,
\end{equation}
so that we encounter a collinear singularity in the function
$G_{ia}(P_{ia}^2,\Delta s)$, but not in $G_{ai}(P_{ia}^2,\Delta s)$.
Therefore, we can obtain $G_{ai}(P_{ia}^2,\Delta s)$ from
Eq.~\eqref{eq:Gai2} upon taking the limit $m_i\to0$, but
to obtain $G_{ia}(P_{ia}^2,\Delta s)$ for small $m_i$
we have to perform a new integration.
The explicit results are
\begin{align}
G_{ia}(P_{ia}^2,\Delta s) &=
{\cal L}\left(\frac{\bar P_{ia}^4}{m_a^2-\bar P_{ia}^2},m_i^2\right)
-\ln^2\left(\frac{\Ds(m_a^2-\bar P_{ia}^2)}{\bar P_{ia}^4}\right)
-\frac{3}{2}\ln\left(\frac{\Ds(m_a^2-\bar P_{ia}^2)}{\bar P_{ia}^4}\right)
-\frac{2\pi^2}{3}
+\frac{3}{2}
\nonumber\\
&\quad {}
+ \theta(-P_{ia}^2)\left[
\ln^2\left(\frac{m_a^2-\bar P_{ia}^2}{-\bar P_{ia}^2}\right)
+2\Li_{2}\left(\frac{-\bar P_{ia}^2}{m_a^2-\bar P_{ia}^2}\right)
-\frac{\pi^2}{6}
\right],
  \label{eq:Gia3}
\\
G_{ai}(P_{ia}^2,\Delta s) &=  
{\cal L}\left(m_a^2-\bar P_{ia}^2,m_a^2\right)
-2\ln\left(\frac{\Ds}{-\bar P_{ia}^2}\right)
\ln\left(\frac{m_a^2}{m_a^2-\bar P_{ia}^2}\right)
-2\ln\left(\frac{\Ds}{-\bar P_{ia}^2}\right)
-\frac{3}{2}\ln\left(\frac{m_a^2}{m_a^2-\bar P_{ia}^2}\right)
\nonumber\\
&\quad {}
+ \theta(-P_{ia}^2) \biggl[
2\Li_{2}\left(\frac{\bar P_{ia}^2}{m_a^2}\right)
+2\ln\left(\frac{m_a^2}{m_a^2-\bar P_{ia}^2}\right)
\ln\left(\frac{m_a^2}{-\bar P_{ia}^2}\right)
+\frac{\pi^2}{6}
+2\ln\left(\frac{m_a^2}{-\bar P_{ia}^2}\right)
+2 
+\frac{2m_a^2}{\bar P_{ia}^2} \biggr].
  \label{eq:Gai3}
\end{align}
In order to calculate
the function $\bar{\cG}_{ia}(P_{ia}^2,\Delta s,z)$ for
the non-collinear-safe contribution 
we have to perform the integration in Eq.~\eqref{eq:cGia} over the $x_{ia}$~range,
which is bounded by
\begin{align}
1-\Delta x + {\cal O}(\Delta x^2) < x_{ia} < 1+\frac{m_i^2(1-z)}{\bar P_{ia}^2 z} +{\cal O}(m_i^2/\bar P_{ia}^2) .
\end{align}
A simple integration yields
\begin{align}
 \bar{\cG}_{ia}(P_{ia}^2,\Delta s,z)= &\, 
\frac{1+z^2}{1-z}\ln\left(\frac{\Ds z}{m_i^2 (1-z)}\right)-\frac{2z}{1-z}.
  \label{eq:cGia2}
\end{align}

\paragraph{Light particle $a$ and massive particle $i$}

If particle $a$ is light and particle $i$ is heavy, 
we have to consider the hierarchy 
\begin{equation}
m_\gamma^2\ll m_a^2 \ll \Ds \ll m_i^2,|P_{ia}^2|,|\bar P_{ia}^2|,
\end{equation}
so that we encounter a collinear singularity in the function
$G_{ai}(P_{ia}^2,\Delta s)$, but not in $G_{ia}(P_{ia}^2,\Delta s)$.
We can obtain $G_{ia}(P_{ia}^2,\Delta s)$ from
Eq.~\eqref{eq:Gia2} upon taking the limit $m_a\to0$, but
to obtain $G_{ai}(P_{ia}^2,\Delta s)$ for small $m_a$
we have to perform another integration.
The explicit results are
\begin{align}
G_{ia}(P_{ia}^2,\Delta s) &=
{\cal L}(m_i^2-\bar P_{ia}^2,m_i^2)
-2\ln\left(\frac{\Ds}{-\bar P_{ia}^2}\right)
\ln\left(\frac{m_i^2}{m_i^2-\bar P_{ia}^2}\right)
-2\ln\left(\frac{\Ds}{-\bar P_{ia}^2}\right)
-\frac{3}{2}\ln\left(\frac{m_i^2}{m_i^2-\bar P_{ia}^2}\right)
\nonumber\\
&\quad {}
+ \theta(-P_{ia}^2)\Biggl[ 
\ln^2\left(\frac{m_i^2}{m_i^2-\bar P_{ia}^2}\right)
+2\Li_{2}\left(\frac{m_i^2}{m_i^2-\bar P_{ia}^2}\right)
-\frac{\pi^2}{6}
+2\ln\left(\frac{m_i^2}{-\bar P_{ia}^2}\right)
+\frac{2m_i^2}{\bar P_{ia}^2}
+2\Biggr],
  \label{eq:Gia4}
\\
G_{ai}(P_{ia}^2,\Delta s) &=   
{\cal L}\left(\frac{\bar P_{ia}^4}{m_i^2-\bar P_{ia}^2},m_a^2\right)
-\ln^2\left(\frac{\Ds(m_i^2-\bar P_{ia}^2)}{\bar P_{ia}^4}\right)
-\frac{3}{2}\ln\left(\frac{\Ds(m_i^2-\bar P_{ia}^2)}{\bar P_{ia}^4}\right)
-\frac{\pi^2}{3}
+\frac{1}{4}
\nonumber\\
&\quad
+ \theta(-P_{ia}^2)\Biggl[ 
\ln^2\left(\frac{m_i^2-\bar P_{ia}^2}{-\bar P_{ia}^2}\right)
+2\Li_{2}\left(\frac{-\bar P_{ia}^2}{m_i^2-\bar P_{ia}^2}\right)
-\frac{\pi^2}{6}
\Biggr].
  \label{eq:Gai4}
\end{align}
In order to calculate
the function ${\cG}_{ai}(P_{ia}^2,\Delta s,x)$ for
the non-collinear-safe contribution 
we have to perform the integration in Eq.~\eqref{eq:cGai} over the $z_{ia}$~range,
which is bounded by
\begin{align}
1-x\Delta x < z_{ia} < 1+\frac{m_a^2}{\bar P_{ia}^2}x(1-x).
\end{align}
A simply integration yields
\begin{align}
 {\cG}_{ai}(P_{ia}^2,\Delta s,x)= &\, 
\frac{1+x^2}{1-x}\ln\left(\frac{\Ds}{m_a^2 (1-x)}\right)-\frac{2x}{1-x}.
  \label{eq:cGai2}
\end{align}

\paragraph{Light particles $a$ and $i$}

Finally, we consider the case of both $a$ and $i$ being light, so that the mass hierarchy is
\begin{equation}
m_\gamma^2\ll m_a^2,m_i^2 \ll \Ds \ll |P_{ia}^2|.
\end{equation}
Here we always have $P_{ia}^2<0$.
The full integrals $G_{ia}(P_{ia}^2,\Delta s)$ and $G_{ai}(P_{ia}^2,\Delta s)$
can be obtained from the results above where the respective emitter is light
upon taking the limit of a massless spectator,
i.e.\ upon taking $m_a\to0$ in Eq.~\eqref{eq:Gia3}
and $m_i\to0$ in Eq.~\eqref{eq:Gai4}.
The results are
\begin{align}
G_{ia}(P_{ia}^2,\Delta s) &=
{\cal L}(-P_{ia}^2,m_i^2)
-\ln^2\left(\frac{\Ds}{-P_{ia}^2}\right)
-\frac{3}{2}\ln\left(\frac{\Ds}{-P_{ia}^2}\right)
-\frac{\pi^2}{2}
+\frac{3}{2},
  \label{eq:Gia5}
\\
G_{ai}(P_{ia}^2,\Delta s) &=  
{\cal L}(-P_{ia}^2,m_a^2)
-\ln^2\left(\frac{\Ds}{-P_{ia}^2}\right)
-\frac{3}{2}\ln\left(\frac{\Ds}{-P_{ia}^2}\right)
-\frac{\pi^2}{6}
+\frac{1}{4}.
  \label{eq:Gai5}
\end{align}
The corresponding functions $\bar{\cG}_{ia}(P_{ia}^2,\Delta s,z)$ and
${\cG}_{ai}(P_{ia}^2,\Delta s,x)$ for non-collinear-safe 
observables are the same as given in Eqs.~\eqref{eq:cGia2}
and \eqref{eq:cGai2}, respectively.

\subsubsection{Initial-state emitter and spectator}
\label{subsect:initialE-initialS}

%
\begin{figure}
  \centering
  \includegraphics[scale=1.25]{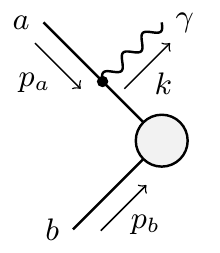}
\caption{Generic diagram with an initial-state emitter $a$ and an initial-state spectator $b$.}
\label{fig:initial-initial}
\end{figure}
Although not needed for the treatment of decays, for completeness here we collect
the formulas with both emitter $a$ and spectator $b$ in the initial state
with masses $m_a$ and $m_b$.
The corresponding structural diagram is shown in Fig.~\ref{fig:initial-initial}. 
Using again the kinematical variables introduced in Ref.~\cite{Dittmaier:1999mb},
\begin{align}
x_{ab} &= \frac{p_a p_b-p_a k-p_b k}{p_a p_b}, \qquad\qquad
y_{ab} = \frac{p_a k}{p_a p_b}
\label{eq:xabyab}
\end{align}
and
\begin{align}
s &= (p_a+p_b)^2, & \bar s &= s-m_a^2-m_b^2, &
\lambda_{ab}&=\lambda(s,m_a^2,m_b^2),
\end{align}
the integrals over the full soft--collinear region is defined as
\begin{equation}
G_{ab}(s,\Ds)
= \int_{x_0}^{x_1}\rd x_{ab}\,
\frac{x_{ab}\bar s^2}{2\sqrt{\lambda_{ab}}}
\int_{y_1(x_{ab})}^{y_2(x_{ab})}\rd y_{ab} \, \gsub{ab}(p_a,p_b,k)
(\bar\theta_a\bar\theta_b + \bar\theta_a\theta_b + \theta_a\bar\theta_b)
\label{eq:Gab}
\end{equation}
with the integration boundary defined by
\begin{align}
y_{1,2}(x) &= \frac{\bar s+2m_a^2}{2s}(1-x) \mp
\frac{\sqrt{\lambda_{ab}}}{2s}
\sqrt{(1-x)^2-\frac{4m_\gamma^2 s}{\bar s^2}}
\nonumber\\
x_1 &= 1-\frac{2m_\gamma \sqrt{s}}{\bar s},
\end{align}
where again $x_0$ is the lowest kineamatically allowed value.
The soft and collinear regions are characterized as follows,
\begin{subequations}
\label{eqs:iniEinitS}
\begin{align}
  \label{eq:iniEinitS_softcond}
  &\text{soft region:} &
  k & \to 0 &
  \begin{cases}
  s_{a\gamma}<\Ds &\Rightarrow\; y_{ab} < \Delta y, \\
  s_{b\gamma}<\Ds &\Rightarrow\; x_{ab} > 1-y_{ab}-\Delta y ,
  \end{cases}
  \\[.5em]
  \label{eq:iniEinitS_coll1cond}
  &\text{$a$-collinear region:} &
  k&\nrightarrow0, \quad k \parallel p_a &
  \begin{cases}
  s_{a\gamma}<\Ds &\Rightarrow\; y_{ab} < \Delta y, \\
  s_{b\gamma}>\Ds &\Rightarrow\; x_{ab} < 1-y_{ab}-\Delta y ,
  \end{cases}
  \\[.5em]
  \label{eq:iniEinitS_coll2cond}
  &\text{$b$-collinear region:} &
  k&\nrightarrow0, \quad k \parallel p_b &
  \begin{cases}
  s_{a\gamma}>\Ds &\Rightarrow\; y_{ab} > \Delta y, \\
  s_{b\gamma}<\Ds &\Rightarrow\; x_{ab} > 1-y_{ab}-\Delta y ,
  \end{cases}
\end{align}
\end{subequations}
where
\begin{equation}
\Delta y=\frac{\Ds}{\bar s}.
\end{equation}
Similar to the previous cases of emitter/spectator configurations,
in addition to the full integral \eqref{eq:Gab} we have to introduce
the partial integral 
\begin{align}
\cG_{ab}(s,\Delta s,x) =
\frac{x\bar s^2}{2\sqrt{\lambda_{ab}}}
\int_{y_1(x)}^{y_2(x)}\rd y_{ab} \, \gsub{ab}(p_a,p_b,k)
(\bar\theta_a\bar\theta_b + \bar\theta_a\theta_b + \theta_a\bar\theta_b)
  \label{eq:cGab}
\end{align}
to keep track of the momentum flow in the collinear region of the emitter.
Again we used $x=x_{ab}$ to shorten the notation.
The momentum $xp_a$ is the fermion momentum after collinear
photon emission from the initial state, i.e.\ the momentum $x p_a$ enters the hard
non-radiative process.
The soft singularities in the integrals again are extracted from the numerical integration upon
introducing plus distributions in the integrals,
\begin{align}
{\cG}_{ab}(s,\Delta s,x) &=
\left[{\cG}_{ab}(s,\Delta s,x)\right]_+
+ \delta(1-x) G_{ab}(s,\Delta s).
\end{align}

\paragraph{Massive initial-state emitter $a$ and spectator $b$}

If both initial-state emitter $a$ and initial-state spectator $b$ 
are massive, we have to respect the following hierarchy of parameters,
\begin{equation}
m_\gamma^2 \ll \Delta s \ll m_a^2,m_b^2,s,\bar s.
\label{eq:massiveabhierarchy}
\end{equation}
No collinear singularities are encountered, and the effective integration
region in $G_{ab}(s,\Delta s)$ is a neighbourhood of the point
$(x_{ab},y_{ab})=(1,0)$, so that the function
${\cG}_{ab}(s,\Delta s,x)$ is not required.
The full integral reads
\begin{align}
G_{ab}(s,\Delta s) &=
\frac{\bar s}{\sqrt{\lambda_{ab}}}\left\{
\ln(d_1)\ln\left(\frac{m_\gamma^2 \lambda_{ab}}{\Ds^2 s}\right)
+2\Li_{2}(d_1)
-\frac{\pi^2}{3}+\frac{1}{2}\ln^2(d_1)
\right\}
+2\ln\left(\frac{m_\gamma m_a}{\Ds}\right)+2 
\nonumber\\
&\quad
+ f_{ab,+}(s) \;\mp\; \theta(2m_a^2+2m_b^2-s)\,f_{ab,\pm}(s)
\qquad \mbox{for} \quad m_a \lessgtr m_b,
  \label{eq:Gab2}
\end{align}
with the auxiliary functions
\begin{align}
f_{ab,\pm}(s)  =
\pm\frac{\bar s}{\sqrt{\lambda_{ab}}}\Biggl\{ &
\ln^2\left(\frac{\bar s+2m_a^2\pm\sqrt{\lambda_{ab}}}{2s}\right)
-2\ln\left(\frac{\bar s+2m_a^2\pm\sqrt{\lambda_{ab}}}{2s}\right)
\ln\left(\frac{\bar s+2m_b^2\mp\sqrt{\lambda_{ab}}}{2s}\right)
\nonumber \\
& {}
-2\Li_{2}\left(\frac{\bar s+2m_b^2\mp\sqrt{\lambda_{ab}}}{2s}\right)
+\frac{\pi^2}{6}
+\frac{\bar s\mp\sqrt{\lambda_{ab}}}{\bar s}
\ln\left(\frac{\bar s+2m_a^2\pm\sqrt{\lambda_{ab}}}{\bar s+2m_b^2\mp\sqrt{\lambda_{ab}}}\right)
\nonumber \\
& {}
+1-\frac{2m_a^2}{\bar s} \mp \frac{\sqrt{\lambda_{ab}}}{\bar s} \Biggr\},
\end{align}
Here we 
partially used the results of Ref.~\cite{Dittmaier:1999mb}, 
where we have identified $(1-x_0)$ with $2\Delta y$
in the limit $\Delta y\rightarrow0$, and 
\begin{equation}
 d_1=\frac{\bar s+2m_a^2-\sqrt{\lambda_{ab}}}{\bar s+2m_a^2+\sqrt{\lambda_{ab}}}.
\label{eq:d_1}
\end{equation}
There is no need to introduce the function
${\cG}_{ab}(s,\Delta s,x)$
if the emitter particle is massive.

\paragraph{Massive emitter $a$ and light spectator $b$}

The hierarchy between the parameters is 
\begin{equation}
m_\gamma^2\ll \Ds \ll m_a^2,s,\bar s, \qquad m_b=0.
\end{equation}
Similar to the previous cases,
the integral $G_{ab}(s,\Delta s)$ can be obtained upon
taking the limit $m_b\to0$ from the result of the previous section,
\begin{align}
G_{ab}(s,\Delta s) &=
{\cal L}(s,m_a^2) 
-2\ln\left(\frac{\Ds}{\bar s}\right)
\ln\left(\frac{m_a^2}{s}\right)
-2\ln\left(\frac{\Ds}{\bar s}\right)
-\frac{3}{2}\ln\left(\frac{m_a^2}{s}\right)
\nonumber\\
&\quad
+ \theta(s-2m_a^2)
\Biggl\{
\ln^2\left(\frac{m_a^2}{s}\right)
+2\Li_{2}\left(\frac{m_a^2}{s}\right)
-\frac{\pi^2}{6}
+2\ln\left(\frac{m_a^2}{\bar s}\right)
+2-\frac{2m_a^2}{\bar s} \Biggr\}.
  \label{eq:Gab3}
\end{align}
Similarly to the case of non-vanishing $m_a$ and $m_b$, 
$G_{ab}(s,\Delta s)$ does not comprise an integral over the
energy flow in the collinear regions, so that the quantity 
${\cG}_{ab}(s,\Delta s,z)$ is not needed either.

\paragraph{Light emitter $a$ and massive spectator $b$}

If the emitter $a$ is light and the spectator $b$ is heavy, 
we have to consider the hierarchy 
\begin{equation}
m_\gamma^2\ll m_a^2 \ll \Ds \ll m_b^2,s,\bar s.
\end{equation}
We encounter a collinear singularity in the function
$G_{ab}(s,\Delta s)$, i.e.\
we have to perform another integration
to obtain $G_{ab}(s,\Delta s)$ for small $m_a$.
The explicit result is
\begin{align}
G_{ab}(s,\Delta s) &=
{\cal L}\left(\frac{\bar s^2}{s},m_a^2\right)
-\ln^2\left(\frac{\Ds s}{\bar s^2}\right)
-\frac{3}{2}\ln\left(\frac{\Ds s}{\bar s^2}\right)
-\frac{\pi^2}{3}
+\frac{1}{4}
\nonumber\\
& \quad
+ \theta(s-2m_b^2)\Biggl[ 
\ln^2\left(\frac{\bar s}{s}\right)
+2\Li_{2}\left(\frac{\bar s}{s}\right)
-\frac{\pi^2}{6}
\Biggr] .
  \label{eq:Gab4}
\end{align}
In order to calculate
the function ${\cG}_{ab}(s,\Delta s,x)$ 
we have to perform the integration in Eq.~\eqref{eq:cGab} over the $y_{ab}$~range,
which is bounded by
\begin{align}
\frac{m_a^2}{\bar s}(1-x) < y_{ab} < \Delta y.
\end{align}
A simple integration yields
\begin{align}
 {\cG}_{ab}(s,\Delta s,x)= &\, 
\frac{1+x^2}{1-x}\ln\left(\frac{\Ds}{m_a^2 (1-x)}\right)-\frac{2x}{1-x}.
  \label{eq:cGab2}
\end{align}

\paragraph{Light emitter $a$ and spectator $b$}

Finally, we consider the case of both $a$ and $b$ being light, so that the mass hierarchy is
\begin{equation}
m_\gamma^2\ll m_a^2,m_b^2 \ll \Ds \ll s.
\end{equation}
The full integral $G_{ab}(s,\Delta s)$
can be obtained from the previous case 
upon taking the limit of a massless spectator,
i.e.\ upon taking $m_b\to0$ in Eq.~\eqref{eq:Gab4}.
The result is
\begin{align}
G_{ab}(s,\Delta s)=&
{\cal L}(s,m_a^2) 
-\ln^2\left(\frac{\Ds}{s}\right)
-\frac{3}{2}\ln\left(\frac{\Ds}{s}\right)
-\frac{\pi^2}{6}
+\frac{1}{4}.
  \label{eq:Gab5}
\end{align}
The corresponding function ${\cG}_{ab}(s,\Delta s,x)$ 
is the same as given in \eqref{eq:cGab2}.


\section{Application to the top-quark decay}
\label{sec:top}

As an application of the techniques described in Sects.~\ref{sec:dipole} and \ref{sec:ocsm}, in the following we present the calculation of the QCD and EW 
radiative corrections to the top-quark decay at NLO.
First, we summarize our calculational setup in Sect.~\ref{sec:top:setup}. Then, we discuss the different contributions to the top-quark decay width in Sect.~\ref{sec:top:total}.
Finally, in Sect.~\ref{sec:top:diff} we present differential distributions in the top-quark rest frame for a set of kinematic observables.

\subsection{Calculational setup and input parameters}
\label{sec:top:setup}

Following Ref.~\cite{Beringer:1900zz}, the set of input parameters used in our numerical evaluation is given by
\begin{align}
  m_{\Pqt} &= 173.34~\GeV, &
  m_{\Pqb} &= 4.78~\GeV,
  \nonumber\\
  M_{\PW}^{\OS} &= 80.385~\GeV, &
  \Gamma_{\PW}^{\OS} &= 2.085~\GeV,
  \nonumber\\
  M_{\PZ}^{\OS} &= 91.1876~\GeV, &
  \Gamma_{\PZ}^{\OS} &= 2.4952~\GeV,
  \nonumber\\
  m_{\Pgm} &= 105.6583715~\MeV, &
  M_{\PH} &= 125.9~\GeV,
  \nonumber\\
  G_{\Pgm} &= 1.1663787 \cdot 10^{-5}~\GeV^{-2}, &
  \alphas(M_{\PZ}) &= 0.119,
  \label{eq:input-parameters}
\end{align}
with the masses $m_{\Pqt}$ and $M_{\PH}$ taken from Refs.~\cite{ATLAS:2014wva} and~\cite{Dittmaier:2014qza}, respectively.
Light fermions, i.e.\ all fermions but the top and bottom quarks, are considered massless. 
Further, we do not consider any mixing among the quark flavours in our calculation, i.e.\ we set the CKM matrix equal to the unit matrix.
This approximation is justified because 
flavour transitions involving the third generation are 
practically forbidden ($V_{\Pqt\Pqb}\approx1$).
Furthermore, any impact of a mixing among the first two generations cancels in the final result after taking the flavour sums owing to the unitarity of the CKM matrix and the approximation of vanishing quark masses for the first two generations.

In order to describe the $\PW$-boson resonance,
we employ the \emph{complex-mass scheme}~\cite{Denner:1999gp, Denner:2005fg, Denner:2006ic}, a method that takes into account the effects of the instability and the off-shellness of the unstable particle in a gauge-invariant way, providing a consistent procedure at the one-loop level. 
The main idea of the complex-mass scheme is to consider the squared boson masses as complex quantities, defined through the gauge-invariant pole mass $M_V$ and width $\Gamma_V$, 
\begin{equation}
  \mu_V^2 = M_V^2- \ri M_V\Gamma_V, \qquad V=\PW,\PZ,
  \label{eq:complMass}
\end{equation}
and to introduce them directly at the level of the Lagrangian density. 
To apply the 
complex-mass scheme, the on-shell masses of the $\PW$ and $\PZ$ bosons given in Eq.~\eqref{eq:input-parameters} are converted to the corresponding pole masses according to \cite{Sirlin:1991fd,Bardin:1988xt,Beenakker:1996kn}
\begin{align}
  M_V &= \frac{M_V^{\OS}}{c_V}, &
  \Gamma_V &= \frac{\Gamma_V^{\OS}}{c_V}, &
  c_V &= \sqrt{1+\left(\frac{\Gamma_V^{\OS}}{M_V^{\OS}}\right)^2}.
\label{eq:MGpole}
\end{align}

To preserve gauge invariance, the complex masses have to be introduced in a consistent manner.
In particular, this implies a complex 
definition of the weak mixing angle, 
\begin{equation}\label{eq:sw2_CMS}
  \sw^2 \equiv 1- \frac{\mu_{\PW}^2}{\mu_{\PZ}^2} ,
\end{equation}
and a generalization of the renormalization procedure for unstable particles.
Further details on the complex-mass scheme can be found in Refs.~\cite{Denner:2005fg, Denner:2006ic}.

The dependence on the bottom-quark mass is fully accounted for without approximation. The value of $m_{\Pqb}$ given in Eq.~\eqref{eq:input-parameters} corresponds to the pole mass through the two-loop conversion formula given in Ref.~\cite{Beringer:1900zz}.
Although all fermions with the exception of the top and bottom quarks are treated as massless in our calculation, a dependence on the 
mass of the light fermion arises when non-collinear-safe observables are considered (see Sect.~\ref{sec:dipole:light+kabelschacht}).
In the following, they are only considered for final-state muons where the experimental discrimination of a collinear muon--photon configuration is possible.
Note that no event selection cuts are applied, so that the issue of collinear safety only arises in the event reconstruction. Thus, only differential distributions are concerned with this issue, while the integrated results, i.e.\ the (partial) decay widths, are collinear safe.

The 
EW coupling constant $\alpha$ is derived from the muon decay constant $G_\mu$ via the relation
\begin{align}
  \alpha_{G_\mu} &= \frac{\sqrt{2} M_{\PW}^2}{\pi}\,G_\mu
  \left(1- \frac{M_{\PW}^2}{M_{\PZ}^2}\right)\, .
\end{align}
This definition corresponds to the so-called $G_\mu$-scheme, which avoids large logarithms induced by the running of $\alpha$
from the Thomson limit ($Q^2=0$) to the 
EW scale ($Q^2\sim M_{\PW}^2$) and further absorbs dominant corrections to the 
$\rho$~parameter into the coupling definition.
Using $\alpha_{G_\mu}$ instead of $\alpha$ in LO predictions changes the charge-renormalization 
constant by a term containing the quantity $\Delta r$~\cite{Sirlin:1980nh,Denner:1991kt}
which quantifies the radiative corrections to muon decay
(see, e.g., Ref.~\cite{Butterworth:2014efa} for further details).
For the strong coupling $\alphas$ we employ the fixed value given in Eq.~\eqref{eq:input-parameters}.

\subsubsection{Details on the calculations}

The implementation employing the slicing method of Section~\ref{sec:ocsm} was performed using matrix elements that were calculated using the Weyl--van-der-Waerden spinor formalism (as formulated in Ref.~\cite{Dittmaier:1998nn}) for the tree-level amplitudes and in-house \texttt{Mathematica} routines for the virtual one-loop corrections.
Optimized phase-space parametrizations, based on Refs.~\cite{Denner:1999gp,Roth:1999kk}, were chosen for each of the pronunced structures in the integrand induced by collinear poles or Breit--Wigner resonances. The numerical integration was performed using the \texttt{Vegas} algorithm~\cite{Lepage:1977sw,Lepage:1980dq}.

A second independent calculation was performed using the subtraction method described in Section~\ref{sec:dipole}.
Here, the matrix elements were obtained with the combined packages \texttt{FeynArts}~\cite{Hahn:2000kx} and \texttt{FormCalc}~\cite{Hahn:1998yk,Nejad:2013ina}.
An independent implementation of routines for the phase-space generation was employed. As in the calculation based on the slicing approach, the numerical integration was performed with  \texttt{Vegas}.

Both implementations resort to the \texttt{Collier} library~\cite{Denner:2014gla}, 
which is mainly based on the results of Refs.~\cite{Denner:2005nn,Denner:2010tr}
for the numerical evaluation of loop integrals and 
supports both dimensional and mass regularization for the treatment of IR
singularities.

\subsection{NLO corrections to the top-quark decay width}
\label{sec:top:total}

In Table~\ref{tab:decay-widths} we collect the numerical results on the various contributions to the top-quark decay width, providing separate results on 
the semi-leptonic and hadronic decay channels.
\begin{table}
\centering
\begin{tabular}{llllll}
\toprule
&& $\Gamma_{\Pqt}^\LO[\mathrm{GeV}]$ & $\Gamma_{\Pqt}^\NLO[\mathrm{GeV}]$ & 
$\delta^{\alphas}[\%]$ & $\delta^\alpha[\%]$ 
\\
\midrule
semi-leptonic & off-shell & $0.161065(1)$ & $0.148109(2)$ & $-9.38$ & $+1.34$
\\
& NWA & $0.163634$ & $0.150504$ & $-9.38$ & $+1.35$
\\
\midrule
hadronic & off-shell & $0.483194(1)$ & $0.46242(3)$ & $-5.59$ & $+1.29$
\\
& NWA & $0.490902$ & $0.46987$ & $-5.59$ & $+1.31$
\\
\midrule
total & off-shell & $1.449582(4)$ & $1.36918(6)$ & $ -6.85$ & $+1.30$
\\
& NWA & $1.472707$ & $1.39126$ & $-6.85$ & $+1.32$
\\
\bottomrule
\end{tabular}
\caption{Contributions to the decay width of the top-quark at NLO in $\alphas$ and $\alpha$, 
divided into the semi-leptonic and hadronic decay channels. Results using the narrow-width approximation 
(NWA) are given for comparison.}
\label{tab:decay-widths}
\end{table}
The relative correction factors $\delta$ are defined with respect to the corresponding LO prediction
$\Gamma_{\Pqt}^\LO$,
\begin{align}
  \delta^{\alphas} &= \frac{ \Gamma_{\Pqt}^{\NLO\,\QCD} - \Gamma_{\Pqt}^\LO }{\Gamma_{\Pqt}^\LO}, &
  \delta^{\alpha} &= \frac{ \Gamma_{\Pqt}^{\NLO\,\EW} - \Gamma_{\Pqt}^\LO }{\Gamma_{\Pqt}^\LO}, 
\end{align}
where $\Gamma_{\Pqt}^{\NLO\,\QCD/\EW}$ denotes the NLO prediction taking into account only QCD/EW
corrections, while the full NLO QCD+EW result is denoted $\Gamma_{\Pqt}^{\NLO}$.
The QCD corrections to the semi-leptonic decay of the top-quark are known in 
semi-analytic form~\cite{Jezabek:1988iv} and are in full agreement with our numerical result.%
\footnote{A readjustment of the prefactor of the result given in Ref.~\cite{Jezabek:1988iv} 
according to $G_\mu \to G_\mu \cdot\Re (\sw^2) / \sw^2$ is required in order to be consistent with the 
complex-mass scheme employed in our calculation.}
The full NLO prediction%
\footnote{The total width is calculated using lepton universality, i.e.\ the $\tau$
mass $m_\tau$ is neglected, since those effects are of ${\cal O}(m_\tau^2/M_\PW^2)$.} 
$\Gamma^{\NLO}_{\Pqt}=1.3693(2)\GeV$ for the total width
(see Table~\ref{tab:decay-widths})
can be compared to the most recent CMS measurement~\cite{Khachatryan:2014nda} of 
\begin{equation}
  \Gamma_{\Pqt}=1.36~\pm0.02~(\text{stat})~{}^{+0.14}_{-0.11}~(\text{syst})~\GeV ,
\end{equation}
assuming $m_{\Pqt}=172.5~\GeV$, revealing good agreement between theory and experiment
at the current accuracy level.

\begin{figure}
  \includegraphics[width=.45\columnwidth]{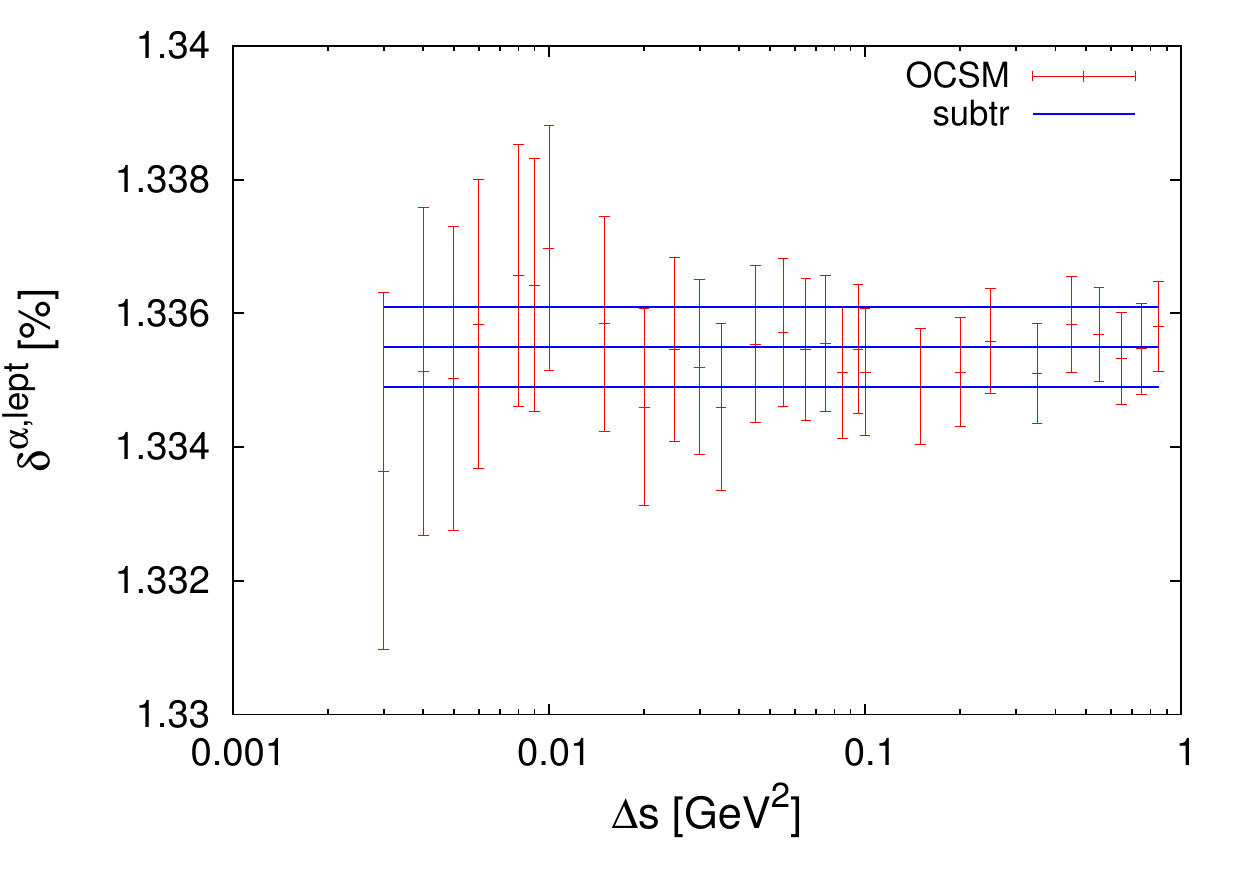}
  \hfill
  \includegraphics[width=.45\columnwidth]{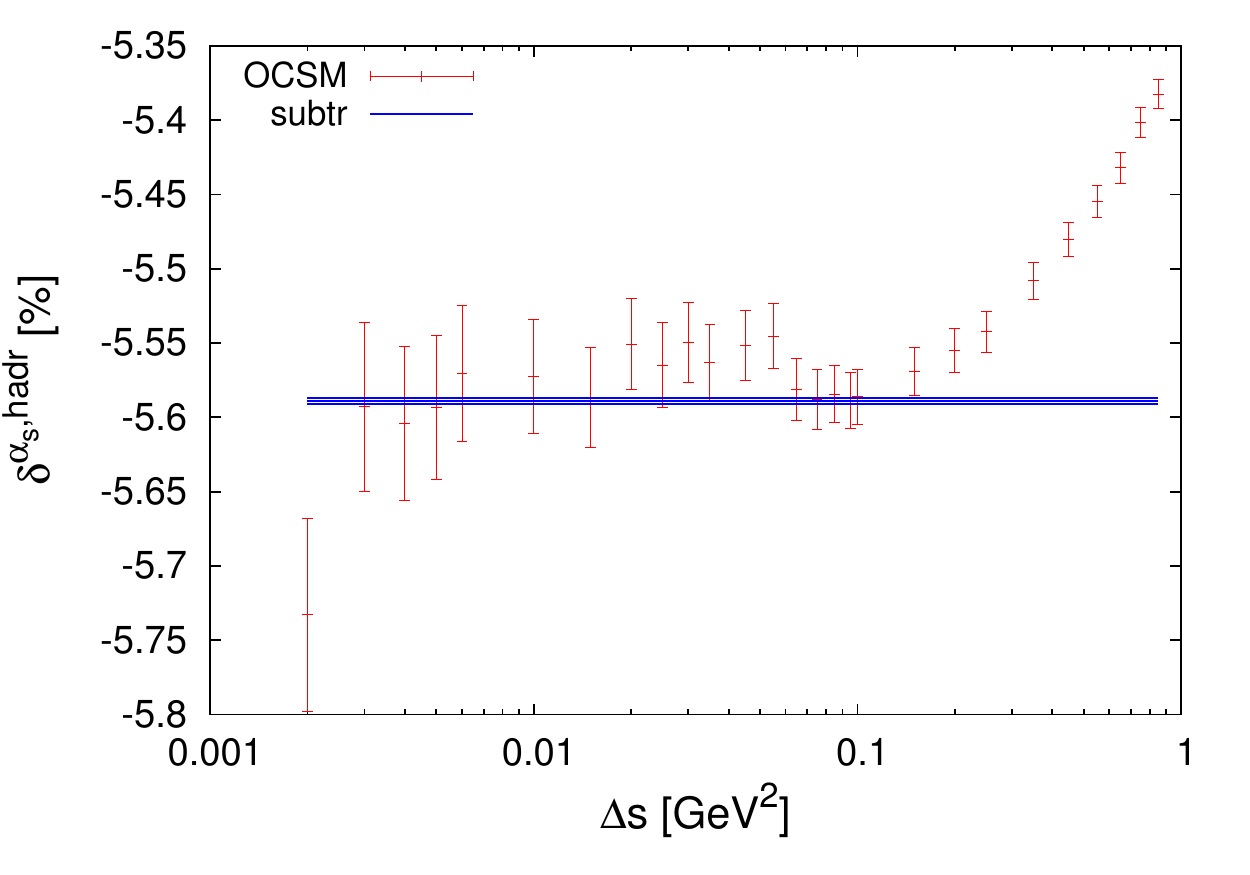}
  \caption{Dependence of the relative correction factor on the cut value $\Ds$ in the OCS method 
(red integration error bars) and comparison to the results obtained using the dipole subtraction method (band indicated by blue lines).
  The upper and lower borders correspond to one standard deviation of the numerical integration.
  The relative EW (left) and QCD (right) corrections are exemplarily 
shown for the case of a top-quark decaying semi-leptonically and hadronically, respectively.
  }
  \label{fig:DeltaS-variation}
\end{figure}

A comparison between the results obtained with
OCS and dipole subtraction is shown in Fig.~\ref{fig:DeltaS-variation}.
As discussed in Sect.~\ref{sec:ocsm}, the slicing approach introduces a cut parameter $\Ds$ in order to isolate the singular region of the real-emission phase space.
The dependence on this cut parameter cancels between the individual parts of the calculation in a non-trivial manner when combined to the final result, constituting 
a powerful check on the correctness of the calculation.
Although this implies a certain degree of arbitrariness in the choice of the specific value for the cut parameter, in practice, sensible values for $\Ds$ are constrained to a certain range.
This interval of validity
is limited by the approximation used in the analytic integration on the upper, and by the stability of the numerical evaluation on the lower end.
It is therefore necessary to perform a scan over the cut parameter to identify the region where both restrictions are simultaneously fulfilled and where the result can be trusted.
Figure~\ref{fig:DeltaS-variation} illustrates such a variation of $\Ds$ covering three orders of magnitude, clearly depicting the on-set of the aforementioned breakdown of the calculation.
Furthermore, we observe good agreement between the results obtained by the two techniques which overlap within their Monte Carlo errors where the $\Ds$ variation develops a plateau.
The OCS results presented in the following are obtained for a specific value of $\Ds$ where a point in the middle of the plateau was chosen after performing such a scan.
This procedure is repeated for each decay mode (semi-leptonic and hadronic) and each type of correction (QCD and EW).

Finally, we have compared the results of our calculation with off-shell $\PW$~bosons to a
narrow-width approximation (NWA) based on on-shell intermediate $\PW$~bosons, which is defined by
\begin{align}
\Gamma^\NWA_{\Pqt\to\Pqb f\bar f'}
&=\Gamma_{\Pqt\to\PW\Pqb}\times{\mathrm{BR}}_{\PW\to f\bar f'}, &
{\mathrm{BR}}_{\PW\to f\bar f'} &= \frac{\Gamma_{\PW\to f\bar f'}}{\Gamma_\PW}
\end{align}
at LO, where ${\mathrm{BR}}_{\PW\to f\bar f'}$ is the
branching ratio of the $\PW$~boson.
Note that we take the experimental result from
Eqs.~\eqref{eq:input-parameters} and \eqref{eq:MGpole} as input
for the total $\PW$~width $\Gamma_\PW$, while the partial widths
are calculated, in order to match the input procedure for the
full top-quark decays with intermediate W~bosons.
At NLO, we linearize the corrections to $\Gamma^\NWA_{\Pqt\to\Pqb f\bar f'}$
in the corrections to the partial decay widths, in order not to spoil
the proper comparison to the results for the full top-quark decays.
Table~\ref{tab:decay-widths} shows that the off-shell result and the NWA for the 
top decay widths differ by about $1.6\%$, 
in good agreement with the naive expectation
of $\order{\Gamma_\PW/M_\PW}$, 
which quantifies typical off-shell effects of the $\PW$~boson in
integrated quantities. The relative corrections are influenced by the off-shell effects
only at the sub per-mille level. 
Off-shell effects of the $\PW$~boson will, however, play
a more important role in differential distributions.

\subsection{Differential distributions in the top-quark rest frame}
\label{sec:top:diff}

In the following, we present differential distributions for the decay of the top-quark in its rest frame.
After describing the event reconstruction in Sect.~\ref{subsect:EvtRec}, we present the numerical results for the semi-leptonic and hadronic decay channels in Sects.~\ref{subsect:ResCorr_lept} and \ref{subsect:ResCorr_hadr}, respectively.
The differential distributions and their corrections are shown as plots that are subdivided into three frames with the following conventions:
The upper panel depicts the absolute distributions, while the bottom two frames show the relative 
EW~(middle panel) and QCD~(bottom panel) corrections.

\subsubsection{Event reconstruction}
\label{subsect:EvtRec}

It is not always possible to fully distinguish the particles of the process in the soft and collinear configurations.
In our calculation, we try to be as close as possible to 
an experimental situation and distinguish the following scenarios:

\paragraph{Electroweak corrections to the semi-leptonic decay}

\begin{description}
\item[``Dresses leptons'':]  
  The EW corrections contain IR singularities originating from soft photons and configurations where the photon becomes collinear to the final-state charged lepton.
  For sufficiently inclusive observables these singularities cancel in the final result.
  To this end, we apply the \emph{photon-recombination procedure} described in Ref.~\cite{Dittmaier:2009cr} which attributes the momentum of a collinear photon to the charged lepton when close to it, 
i.e., when $\Delta R(\Pl,\,\Pgg)=\sqrt{\Delta y(\Pl,\,\Pgg)^2 + \Delta\phi (\Pl,\,\Pgg)^2} < 0.1$, where $\Delta y\,(\Delta\phi)$ is the rapidity (azimuthal angle) difference between two objects.
  Such a recombination procedure is mandatory for a realistic experimental description in the case of electrons.
\item[``Bare muons'':] 
  In the case of muons, a collinear muon--photon pair can be experimentally resolved, so that no photon recombination is required.
  This results in non-collinear-safe observables, which are treated following the procedures described in Sections~\ref{sec:dipole:light+kabelschacht} and \ref{sect:OCSM_IR}.
\end{description}

\paragraph{QCD corrections to the semi-leptonic and hadronic decays}

In contrast to the semi-leptonic case, where we assume
that the bottom quark could be directly identified with a bottom jet, 
in the hadronic case we are confronted with the
situation that experimentally indistinguishable partons are present in the final state.
It is therefore necessary to define an 
IR-safe observable which groups soft and/or collinear partons into jets.
In our calculation we adopt the inclusive generalized anti-$k_\rT$ algorithm~\cite{Cacciari:2008gp,Gerwick:2012fw} for two values of the clustering parameter, $R=0.6$ and $R=1.0$.
The first value corresponds to the ATLAS default choice~\cite{Aad:2011he}, the second one will be motivated in Sect.~\ref{subsect:ResCorr_hadr}.
We assume a bottom-tagging efficiency of $100\%$ and identify the jet containing the $\Pqb$~quark as the $\Pqb$-tagged jet.
In the following, we refer to the jet candidate with the highest transverse momentum%
\footnote{Note that no direction is distinguished in the top-quark rest frame and we have fixed an arbitrary reference axis to compute the transverse component.
}
as the leading jet.

\paragraph{Electroweak corrections to the hadronic decay}

When computing the EW corrections to the hadronic decay, 
we do not distinguish the photon from the other QCD partons and apply the same jet algorithm as described in the previous paragraph.

\subsubsection{Semi-leptonic decay of the top-quark}
\label{subsect:ResCorr_lept}

\begin{figure}
  \includegraphics[width=.45\columnwidth]{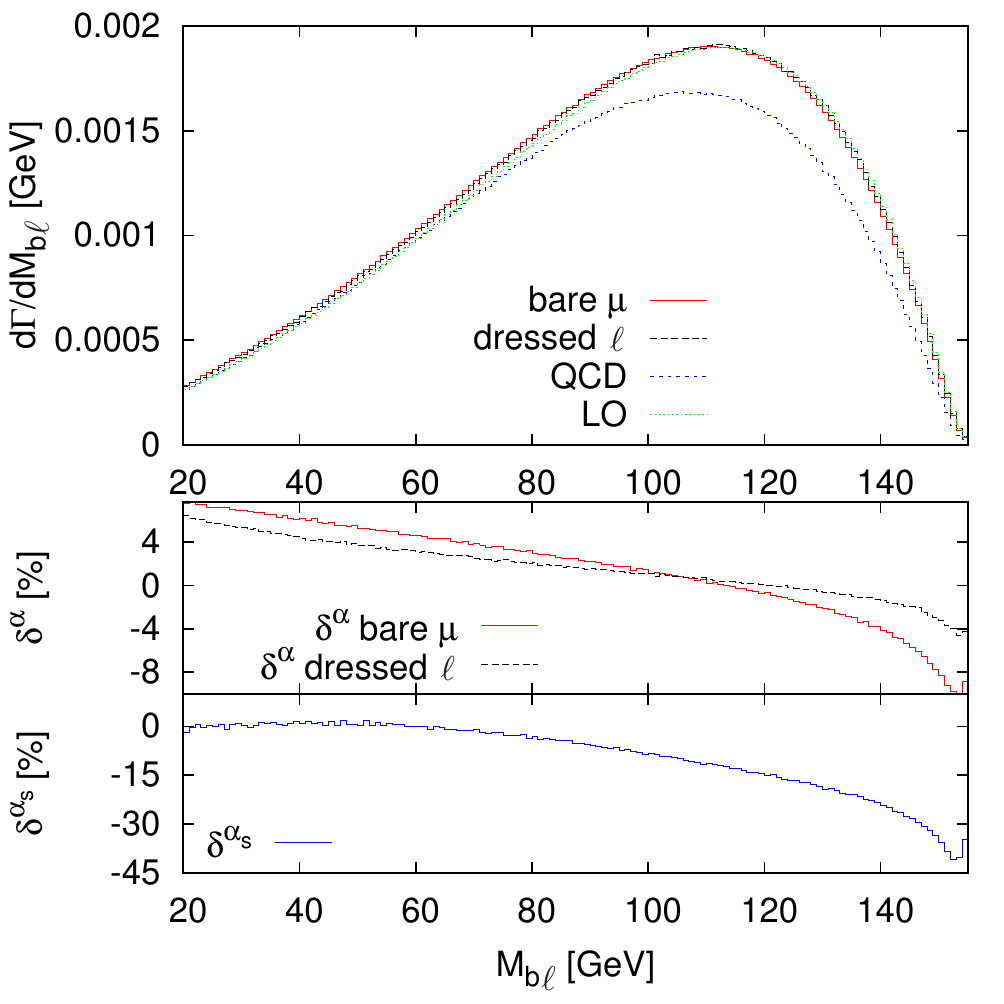}
  \hfill
  \includegraphics[width=.45\columnwidth]{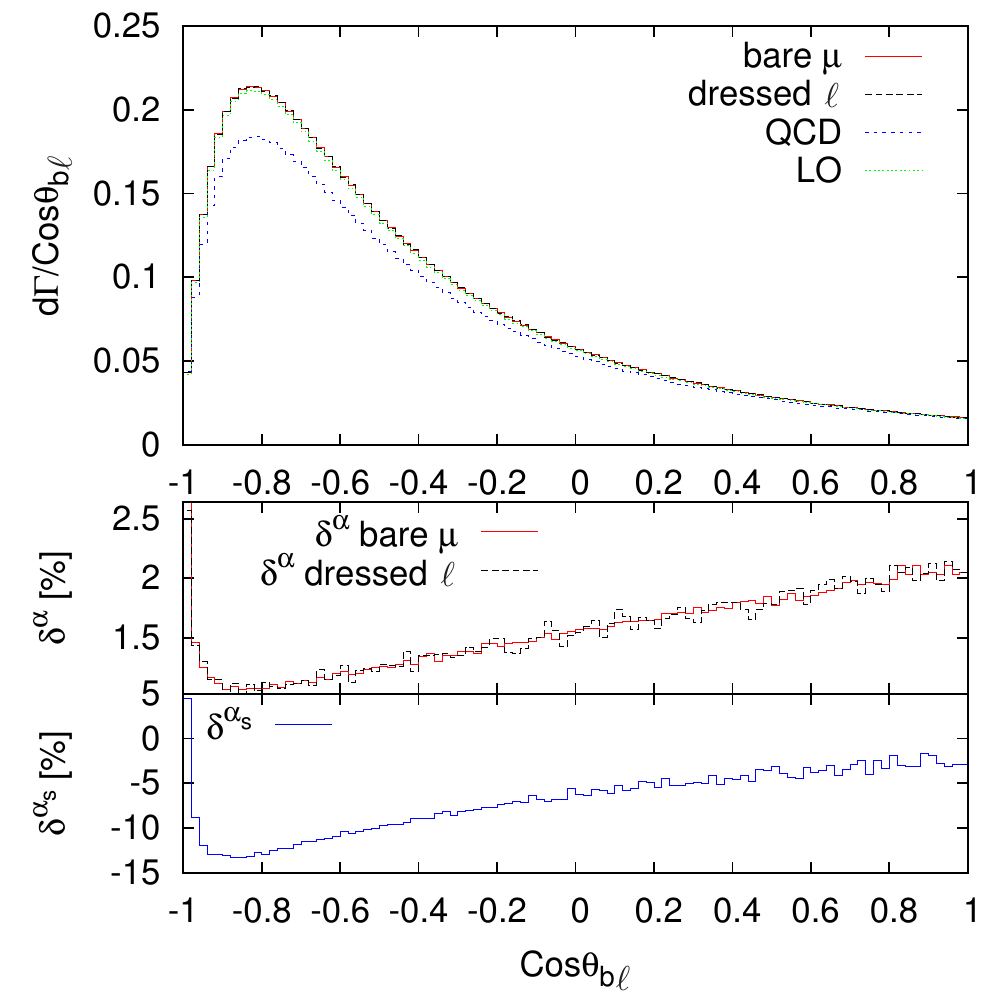}
  \\[1em]
  \includegraphics[width=.45\columnwidth]{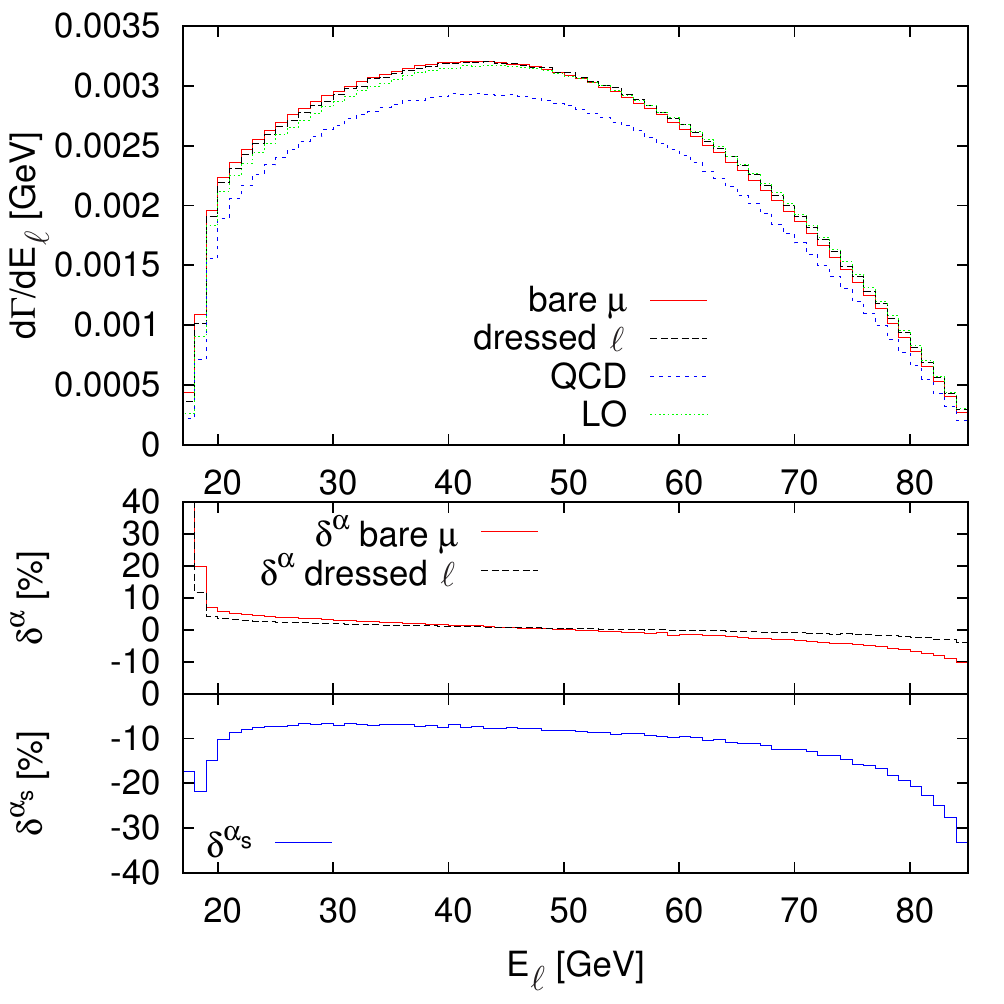}
  \hfill
  \includegraphics[width=.45\columnwidth]{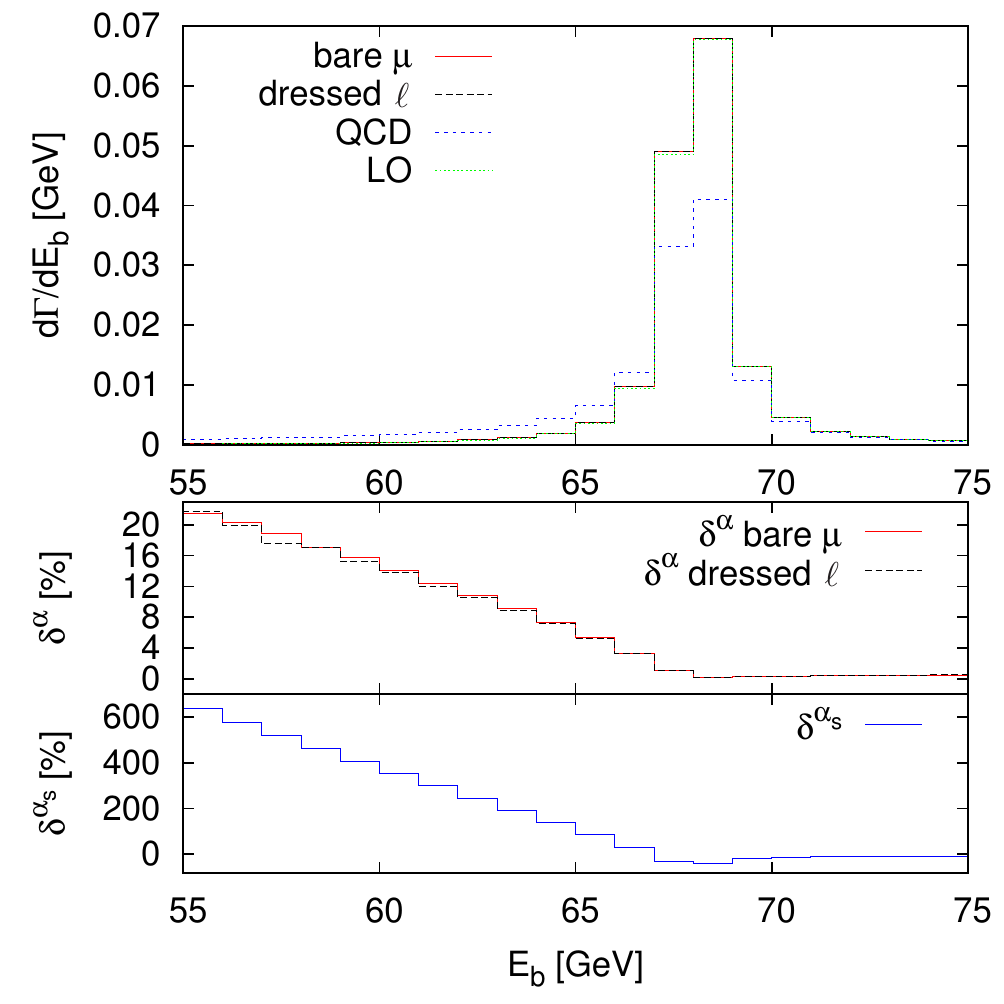}
  \caption{EW and QCD corrections to the semi-leptonic top-quark decay as a function of $M_{\Pqb\ell}$ (top left), $\cos{\theta_{\Pqb\ell}}$ (top right), $E_\Pl$ (bottom left), and $E_\Pqb$ (bottom right), for dressed leptons and bare muons.}
  \label{fig:DistrLep}
\end{figure}

The numerical results for the NLO corrections to the semi-leptonic decay of the top-quark, $\Pqt\to\Pqb\,\Pgnl\,\Plp$, are summarized in Fig.~\ref{fig:DistrLep}.
They comprise differential distributions in the energy of the charged lepton~($E_{\Pl}$) and of the bottom quark~($E_{\Pqb}$), the invariant mass of the bottom quark
and the charged lepton~($M_{\Pqb\Pl}$), and the cosine of the angle between the directions of the lepton and the bottom quark ($\cos{\theta_{\Pqb\Pl}}$).

Following the reconstruction described in Sect.~\ref{subsect:EvtRec}, we distinguish the two cases of ``dressed leptons'' and ``bare muons'' for the EW corrections.
For observables that are not sensitive to the
collinear $\Pl\to\Pl\,\Pgg$ splitting, i.e.\ which are collinear safe, both corrections must be nearly identical.
This can be seen in the distributions in $\cos{\theta_{\Pqb\Pl}}$ and $E_{\Pqb}$ shown in the upper right and lower right plots in Fig.~\ref{fig:DistrLep}, respectively. In the former case in particular, the photon is indeed recombined only if collinear to the positron, i.e.\ it does 
hardly change the direction of the charged lepton. The EW corrections have 
a minimum in the proximity of the peak ($\cos\theta_{\Pqb\ell}\sim-0.8$), are positive and 
smaller than $2\%$.
For comparison, the QCD corrections reach down \mbox{to $-12\%$}.
The distribution in the \Pqb-jet energy is characterized by the fact that
a $1\to2$ particle decay is effectively 
taking place when the $\PW$~boson is on shell. 
In this case, the energy $E_\Pqb$
of the \Pqb-jet tends to the fixed value
\begin{equation}
 E^\OS_\Pqb=\frac{m_\Pqt^2-M_\PW^2+m_\Pqb^2}{2m_\Pqt}.
\end{equation}
This explains the peak at $E_\Pqb\approx 68~\GeV$.
The asymmetric shape of the EW corrections is due to the typical radiative tail produced
by photon radiation off a particle near a resonance energy.
Most $\Pqb$~quarks have an energy $E_\Pqb\approx E^\OS_\Pqb$ after the top-quark decay, 
so that photon radiation off $\Pqb$~quarks, which is collinearly enhanced, reduces
$E_\Pqb$, leading to large positive corrections at energies below $E^\OS_\Pqb$.
This effect is further enhanced in the relative corrections, because the LO distribution
is strongly suppressed there.
The same effect occurs in the QCD correction as well, leading to a similar shape, but
even more than an order of magnitude larger.
Note that in this case corrections more than $100\%$ do not signal a breakdown of perturbation
theory, they merely signal that the effective lowest order is provided by the radiative process
with hard gluon (and partially photon) emission. A proper description of this tail, thus,
would require the inclusion of multi-jet (and multi-photon) emission, as provided, e.g.,
by parton showers.

For the remaining distributions, i.e.\ in
$M_{\Pqb\Pl}$ and $E_{\Pl}$, we observe EW
corrections that are larger in magnitude for ``bare muons'', stemming from the large mass-singular logarithm 
$\alpha\ln m_\mu$
and reflecting the non-collinear-safe nature of the corresponding observable. The emitted photon in the $\Pl\to\Pl\,\Pgg$ splitting removes some energy from the charged lepton. Contrary to the ``bare muon'' case, the ``dressed lepton'' effectively can reabsorb some photons in its redefinition when the photon recombination procedure is applied. Hence, the EW correction to $E_{\Pl}$, always monotonically decreasing, are steeper for the ``bare muon'' than for the ``dressed lepton'', as can be seen in the bottom left plot of Fig.~\ref{fig:DistrLep}.
For similar reasons, this effect is present also in the distribution in $M_{\Pqb\Pl}$, as shown in the top left 
plot of Fig.~\ref{fig:DistrLep}. Here, the EW corrections decrease from $+6\%$ to $-4\%$
for ``dressed leptons'', while they 
drop from $+8\%$ to $-10\%$
when ``bare muons'' are considered in the top-quark decay. The QCD corrections to this distribution are 
decreasing down to $-45\%$ for large invariant masses.

\subsubsection{Hadronic decay of the top-quark}
\label{subsect:ResCorr_hadr}

Here, we present the results for the QCD and EW corrections to the top-quark hadronic decay width.
In our setup, the corrections to the two processes $\Pqt\to \Pqb\, \Pqu \Paqd$ and $\Pqt\to \Pqb\, \Pqc \Paqs$ coincide and are given in Table~\ref{tab:decay-widths} for the integrated decay width.

\begin{figure}
  \includegraphics[width=.45\columnwidth]{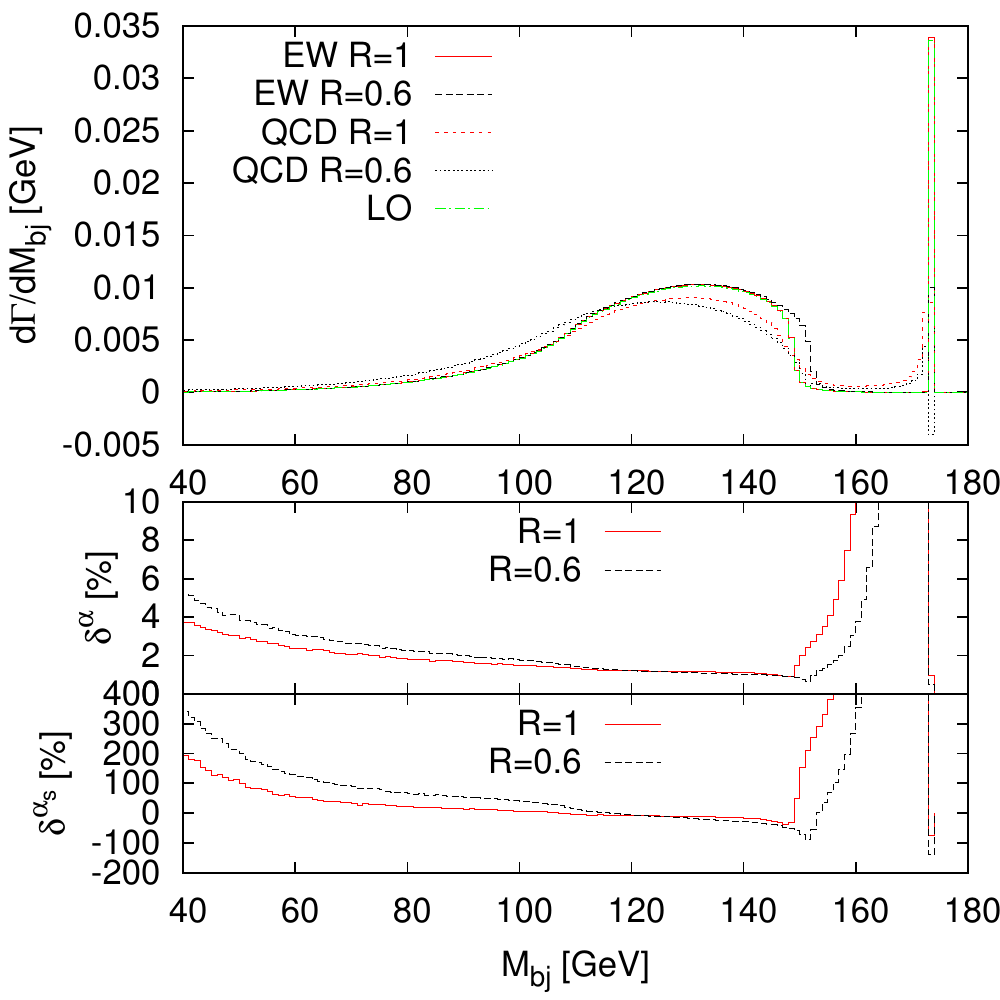}
  \hfill
  \includegraphics[width=.45\columnwidth]{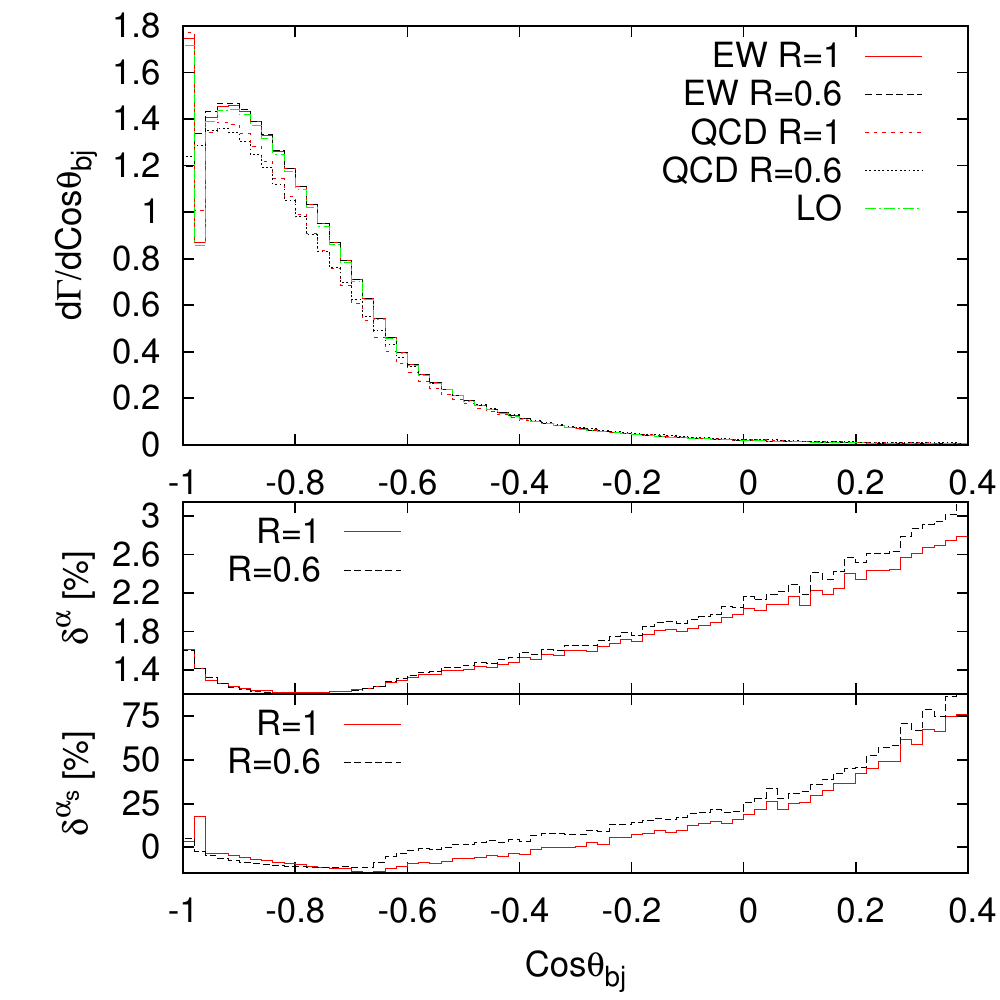}
  \\[1em]
  \includegraphics[width=.45\columnwidth]{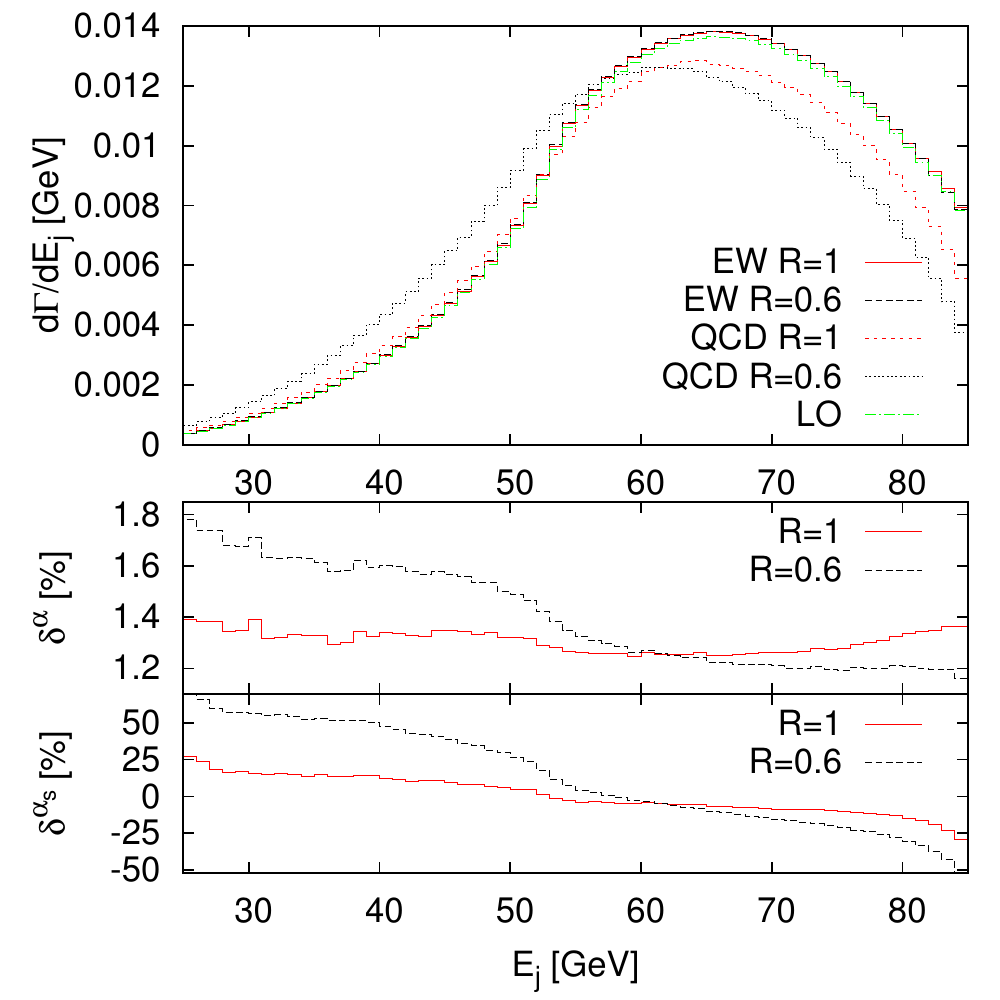}
  \hfill
  \includegraphics[width=.45\columnwidth]{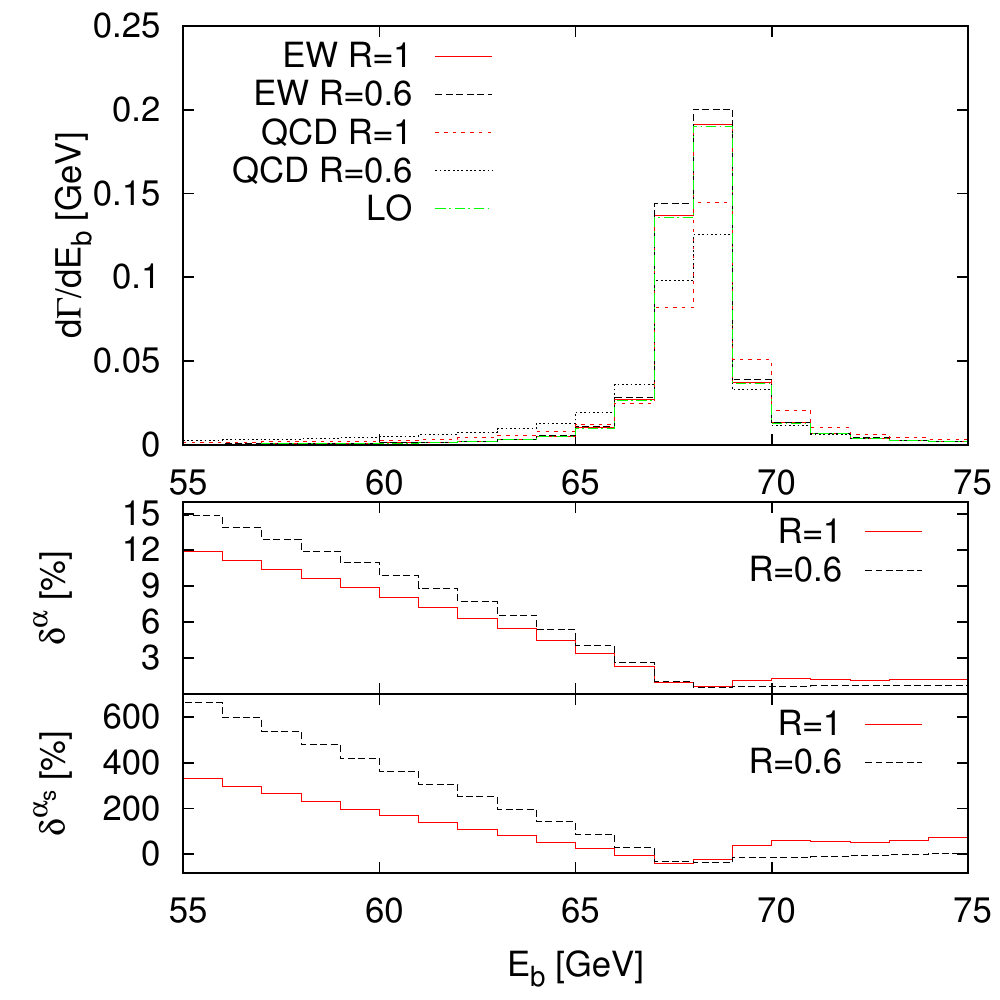}
  \caption{EW and QCD corrections to the hadronic top-quark decay as a function of $M_{\Pqb j}$ (top left), $\cos{\theta_{\Pqb j}}$ (top right), $E_j$ (bottom left), and $E_\Pqb$ (bottom right), with $j$ denoting the leading (non \Pqb-tagged) jet.
  The corrections are shown for the two choices $R=0.6$ and $R=1$ for the angular resolution parameter of 
the jet algorithm.}
  \label{fig:DistrHadr}
\end{figure}

In the top-left plot of Fig.~\ref{fig:DistrHadr} we show the distribution in the invariant mass of the \Pqb-jet and the leading jet.
In the upper frame we notice that the QCD corrections evaluated for $R=0.6$ turn the distribution negative at
$M_{\Pqb j}\sim m_{\Pqt}$. This is of course an unphysical effect:
For a too low value of $R$, as in this case, the recombination between the real and the virtual parts is not sufficiently inclusive, so that
the IR divergence leaves a trace as $\alpha_s\ln R$ corrections.
This problem can be solved by enlarging the cone, e.g.\ to $R=1$, as shown in the same plot.
With a wider cone, there is indeed a larger probability for the particles to be merged. In the case where all the final-state
particles (excluding the \Pqb quark) are combined into a single jet, the invariant mass $M_{\Pqb j}$ reproduces the top-quark mass
value. This explains the appearance of a peak at $M_{\Pqb j}=m_{\Pqt}$.
This obviously happens more frequently for
a $1\to3$ kinematics (as the LO) rather than at NLO. At NLO, the extra
gluon can populate $M_{\Pqb j}$ regions less probable at LO.
This in turns explains the huge corrections in the lower frame of Fig.~\ref{fig:DistrHadr} (top left): 
The magnitude of the NLO correction, 
shown in the middle frame, gets very large due to the strong suppression of the LO in the normalization. 
The EW correction is much smaller than the QCD correction, but is similar in shape.
The curve for $R=0.6$ does not suffer from the same issue as the QCD counterpart. The $R=1$ case is nonetheless shown for comparison.

The distributions obtained choosing $R=0.6$ or $R=1.0$ are compared also for the other observables.
The main difference is that the relative corrections induced 
by the choice of the larger radius are milder than those obtained when using $R=0.6$,
as expected from the inclusiveness argument.
This effect can be clearly seen in the bottom-left and bottom-right plots in Fig.~\ref{fig:DistrHadr}, where the NLO corrections to the energy
of the leading jet and of the \Pqb-jet are presented, respectively.
The energy of the leading jet receives positive QCD corrections
due to gluon radiation, which turn negative above the energy peak at $\sim 65 ~\GeV$.
As a consequence, the shape of the energy distribution at NLO is shifted to the left, more evidently for $R=0.6$. 
In the low-energy tails the relative corrections are enhanced owing the normalization to the small LO prediction in this region. The QCD and EW corrections to the distribution in the energy $E_\Pqb$ of the \Pqb-jet, 
displayed in Fig.~\ref{fig:DistrHadr} (bottom right), again show the distinctive radiative tails
as already discussed for the semi-leptonic top decay in the previous section.

The different choice of the jet-algorithm parameter affects also the
the cosine of the angle between the \Pqb-jet and the leading jet, as
displayed in the top-right plot in Fig.~\ref{fig:DistrHadr}. Once again, a larger jet radius implies a larger probability to merge all final-state partons
other than the \Pqb quark into a single jet. The latter will then be back-to-back to the \Pqb-jet since in this calculation the top-quark is considered at rest. This explains the peak appearing in the bin close to $\cos\theta_{\Pqb j}=-1$ when $R=1$ is employed.
As in the previous case, the shape of the EW corrections match closely the QCD ones,
however, are smaller in size by an order of magnitude.
The strong rise of the QCD corrections toward larger values of $\cos\theta_{\Pqb\mathrm{j}}$ again
originates from the fact that gluon radiation populates the jet phase space more uniformly than at LO,
where there is a strong tendency that the b~quark and the hardest jet are back-to-back.


\section{Conclusions}
\label{sec:concl}

In this paper we have described extensions of two techniques for the treatment of infrared singularities in the computation of electroweak corrections at next-to-leading order:
In Sect.~\ref{sec:dipole} we have supplemented the construction of universal subtraction terms within the dipole subtraction formalism to cover decay kinematics. 
The one-cutoff slicing method described in Sect.~\ref{sec:ocsm} was derived on the basis of the dipole subtraction formalism and organizes the singular contributions in terms of pairs of fermions.
These results represent the first 
ones employing mass regularization and further allow for an arbitrary number of charged 
particles, a case that has not been discussed in detail in the literature so far.
When dealing with more than three charged particles in the process, 
the appearance of new ``surface'' terms has been emphasized.
For both methods we have covered all possible cases of massless and massive fermions in the final state and further worked out the respective extensions for the treatment of non-collinear-safe observables.
The extension of the described techniques to the calculation of NLO QCD corrections is straightforward.

Both methods have been applied to compute the NLO QCD and electroweak corrections to the top-quark decay width.
In this computation, we have
considered the $\Pqb$-quark as massive and evaluated the correction to the top-quark three-body decays fully accounting for the off-shellness of the intermediate $\PW$ boson within the complex-mass scheme. 
Both the hadronic and semi-leptonic decay channels are considered, further accounting for the case of non-collinear-safe observables in the latter. 
For the semi-leptonic decay, the NLO QCD corrections to the partial decay width amount to $\sim -9.4\%$ and $\sim +1.3\%$ for the electroweak corrections.
In the hadronic decay channel, the NLO QCD correction are more moderate with $\sim -5.6\%$, however, still much larger than the EW correction which amount to $\sim +1.3\%$.
While off-shell effects of the W~boson modify the integrated decay widths by about $1.5\%$, the
relative NLO QCD and EW corrections are modified only by few per mille by W~off-shell effects.
We have further presented differential distributions in the top-quark rest frame for different kinematic observables.

The final result for the total decay width of the top-quark at NLO QCD+EW is given 
by $\Gamma_\Pqt^{\NLO} = 1.369~\GeV$, 
which corresponds to an overall correction with respect to the LO prediction of $\delta^{\NLO} = -5.5\%$.
This breaks down to an overall NLO QCD correction of $\sim -6.8\%$, 
and to a NLO EW correction of $\sim +1.3 \%$, normalized to LO.
As mentioned in the introduction, the total top decay width is known at NNLO QCD accuracy.
The $\order{\alphas^2}$ 
correction, evaluated with a finite $\Pqb$-quark mass and a finite $\PW$-boson width 
in a soft-collinear effective theory and normalized to the LO, amounts to a further $\sim -2.1 \%$~\cite{Czarnecki:1998qc,Gao:2012ja}. 
A comparison of our NLO QCD+EW prediction---with and without the additional NNLO correction of
Refs.~\cite{Czarnecki:1998qc,Gao:2012ja}---reveals good agreement with the most
recent measurement by the CMS experiment~\cite{Khachatryan:2014nda}:
\begin{align}
\Gamma^{\mathrm{CMS}}_{\Pqt} &= 1.36\ \pm\ 0.02~(\text{stat})~{}^{+0.14}_{-0.11}~(\text{syst})~\GeV, 
\nonumber\\
\Gamma_\Pqt^{\NLO} &=1.369~\GeV,
\nonumber\\
\Gamma_\Pqt^{\NLO+\NNLO~\QCD} &=1.338~\GeV.
\end{align}
However, the quoted experimental error is still one order of magnitude larger than the typical EW corrections of $\sim\ +1.3\%$ addressed here.
The experimental
error is dominated by systematic uncertainties, such as the
jet energy scale, where a significant improvement is difficult.
The study of top-quark properties is an important part of the physics programme of future lepton colliders which aims for a precision at the order of $10{-}30~\MeV$ in the experimental determination of the top width~\cite{Seidel:2013sqa,Gomez-Ceballos:2013zzn} 
and where the EW corrections are important.

The codes developed in the context of this work can serve as a first step towards an 
unweighted Monte Carlo generator for the decay of the top-quark at NLO,
which can be easily linked to any top-quark production process in order to simulate the decay at NLO accuracy. This is particularly important for processes with 
many particles in the final state when all intermediate particles are decayed. For instance, $\Pqt\overline{\Pqt}+V(V)$ ($V=\Pgg, \PZ, \PWpm$),  
$\Pqt\overline{\Pqt}\PH$, $\Pqt\overline{\Pqt}\Pqb\overline{\Pqb}$, and $\Pqt\overline{\Pqt}\Pqt\overline{\Pqt}$, among others.
Moreover, the top-quark decay width including both NLO QCD and EW corrections and W-boson off-shell effects
comprises an ingredient in NLO predictions for the production of single top-quarks or $\Pqt\bar\Pqt$
pairs including top-quark decays and off-shell effects, such as the process
$\Pp\Pp\to\Pqb\Paqb\PWp\PWm\to6\Pf$, for which NLO QCD corrections 
are known, but not yet the EW corrections.

\appendix
\section*{Appendix}


\section{Conversion from mass regularization to dimensional regularization}
\label{sec:conv}

In this appendix we briefly describe how to convert the results obtained using 
mass regularization (the default choice in the main part of this paper) to the 
corresponding expressions using dimensional regularization in $D=4-2\epsilon$ dimensions.
To this end, we use the results of Refs.~\cite{Catani:2000ef,Catani:2002hc} for the general singular 
IR behaviour of one-loop amplitudes which also include the finite terms that arise from 
the non-commutativity of the limits $\epsilon\to0$ and $m\to0$ and, thus, can be exploited 
for the conversion between the two regularization schemes.
The following translation rules apply both to the functions
$G_{ff'}(P^2,\Ds)$ of the OCS method given in Section~\ref{sec:ocsm} and
to the dipole subtraction functions $D^{(\mathrm{sub})}$ and $G^{(\mathrm{sub})}$
given in Section~\ref{sec:dipole} and Refs.~\cite{Dittmaier:1999mb,Dittmaier:2008md},
respectively. In the following we generically denote these functions $G_{ff'}$.

For an emitter particle with a non-vanishing mass $m$, i.e.\ $m^2\gg\Ds$ in the OCS method,
only soft singularities arise which appear as $\ln m_{\gamma}$ terms in mass regularization.
The transition to dimensional regularization is simply accomplished by the substitution
\begin{align}
\ln(m_{\gamma}^2) \;\to\;
\frac{c_\epsilon\mu^{2\epsilon}}{\epsilon}
+ {\cal O}(\epsilon),
\end{align}
where we have defined the constant
\begin{align}
c_\epsilon = (4\pi)^\epsilon\Gamma(1+\epsilon) 
= \frac{(4\pi)^\epsilon}{\Gamma(1-\epsilon)}
+ \frac{\pi^2}{6}\epsilon^2
+ {\cal O}(\epsilon^3)
\end{align}
and $\mu$ is the reference mass of dimensional regularization.

If the emitter mass $m$ is not light, but is used only as regulator in the limit
$m\to0$ ($m^2\ll\Ds$ in the OCS method), additional collinear singularities
show up as $\ln m$ terms in mass regularization.
The transition from mass to dimensional regularization is more conveniently
described exploiting the fact that only the symmetrized sum $G_{ff'}+G_{f'f}$
of the functions $G_{ff'}$ appears in applications, since each charged particle can appear
as emitter or spectator. In this symmetrized form the results of
Refs.~\cite{Catani:2000ef,Catani:2002hc} can be exploited to derive the
correspondence. However, we have checked that the following rule applies
to each $G_{ff'}$ separately as well.

In mass regularization the IR divergences are always 
contained in the function ${\cal L}(P^2,m^2)$ defined in Eq.~\eqref{eq:L}
with an appropriate kinematical invariant $P^2$. 
In dimensional regularization with a vanishing emitter mass $m=0$ this singular function 
is universally given by
\begin{align}
{\cal L}(P^2,0) =
\left(\frac{\mu^2}{P^2}\right)^\epsilon c_\epsilon \left( \frac{1}{\epsilon^2}
+\frac{3}{2\epsilon} \right) + 2.
\end{align}
Note that the difference between the results in the two regularization schemes,
which is proportional to the combination ${\cal L}(P^2,0)-{\cal L}(P^2,m^2)$,
does not depend on the invariant $P^2$, since the discontinuities arise
from terms proportional to $\left(\frac{m^2}{\mu^2}\right)^{-\epsilon}\frac{1}{\epsilon}$ or
$\left(\frac{m^2}{\mu^2}\right)^{-\epsilon}\frac{1}{\epsilon^2}$, which
vanish in dimensional regulariation ($m\to0$ first), 
but not for mass regulzrization ($\epsilon\to0$ first).


\section*{Acknowledgement}
L.B.\ thanks Patrick Motylinski for some technical help in the phase-space integration.
L.B.\ received support from the Theorie-LHC France initiative of the CNRS/IN2P3 and by the French ANR 12 JS05 002 01 BATS@LHC. 
S.D., L.B.\ and L.O.\ were supported by the Deutsche Forschungsgemeinschaft through the Research Training Group grant GRK\,1102 \textit{Physics at Hadron Accelerators}.
S.D., A.H.\ and L.O.\ received support from the German Research Foundation (DFG) via grant DI 784/2-1. 
Moreover, A.H.\ is supported via the ERC Advanced Grant MC@NNLO (340983).

\bibliographystyle{tep}
\bibliography{topwidth}

\end{document}